\documentclass[12pt]{article}
\usepackage[T1]{fontenc}
\usepackage[utf8]{inputenc}
\usepackage{amsfonts,amsmath,amssymb}
\usepackage{fancyhdr}
\usepackage{array}
\usepackage{graphicx}
\usepackage{fancyhdr}
\usepackage{float}
\usepackage{subfig} % pour inclure les figures
\usepackage[top=25.4mm,bottom=25.4mm,right=25.4mm,left=25.4mm]{geometry}
\begin{document}
%\begin{center}
% \textbf{FILE NOT FOR REVIEW\\}
% \textbf{International Journal for Computational Methods in Engineering Science and Mechanics}
%\end{center}

\begin{center}
	\textbf{ \Large A volume penalization lattice Boltzmann method for simulating flows in the presence of obstacles} 
\end{center} \vspace{-0.1cm}
\begin{center}
	\large $\textrm{M. Benamour}^*$, $\textrm{E. Liberge}^{**}$, $\textrm{C. Béghein}^{**}$
\end{center}  \vspace{-0.5cm}
\begin{center}
  \small (*) CESI, 93 boulevard de la Seine BP602, 92006 Nanterre, France\\
  \small (**) Laboratoire des Sciences de l'Ingénieur pour l'Environnement, UMR 7356 CNRS,\\ Université de La Rochelle, Avenue Michel Crépeau, La Rochelle Cedex 1, France
\end{center} \vspace{0.5cm}

\hspace{-0.6cm}\textbf{\large Abstract}

The LBM is combined with the Volume Penalization (VP LBM) approach to simulate flows in the presence of obstacles. The single relaxation time LBM and 
the multiple relaxation time LBM are used. For cases where the fluid motion is enhanced by moving boundaries, the comparison of our results with 
analytical results is satisfactory. The flow around a cylinder at a relatively high Reynolds number is also calculated. For that case, the VP LBM 
method was parallelized. A comparison of our results with those found in the literature, demonstrates the reliability of the VP LBM to treat complex flows.

\hspace{-0.6cm}\textbf{\small Keywords}: Lattice Boltzmann method, Volume penalization, Navier–Stokes equations.
\section{Introduction}
The approach proposed in this paper consists in coupling the Lattice Boltzmann Method (LBM) with the Volume Penalization (VP) technique 
to model fluid flows in the presence of obstacles. The lattice Boltzmann method, which is easy to parallelize and enables a drastic decrease 
in computing time (in comparison with classical methods: finite differences, finite volumes, finite elements, ...), has been widely used for 
fluid flow simulation \cite{Chen_Doolen_1998, Guo_Shu_2013}. 
In literature, one can notice that various approaches have been implemented in a lattice Boltzmann framework to model flows around
obstacles.\\
In 1994, Ladd \cite{Ladd_1994} adapted the bounce back method to the computation of flows around moving particles, and he proposed
the momentum exchange approach to evaluate the hydrodynamic force acting on the particle. To treat curved boundaries, other authors improved
this method by introducing interpolation formula into the bounce back rule \cite{Bouzidi_2001, Filippova_1998, Mei_1999}. 
Noble and Torczynski \cite{Noble_1998} 
modified the lattice Boltzmann equation, by introducing a source term in the collision term. With this method, the Single Relaxation Time 
Lattice Boltzmann equation could be solved
both in the fluid region, and in regions occupied by a solid obstacle (where the Zou and He bounce back rule \cite{Zou_1997} 
was applied thanks to this source term).\\
The immersed boundary method becomes more and more popular in the lattice Boltzmann community \cite{Wu_2010, Favier_2014, Delouei_2016, Huang_2014, DeRosis_2014}. It consists in solving the Navier-Stokes equations with a source term that mimics the influence of the boundary on the 
flow, on a fixed Eulerian mesh. The
solid boundary (the immersed boundary) is tracked using a Lagrangian approach \cite{Peskin_2002}. Feng and Michaelides \cite{Feng1} and Ten 
Cate et al. \cite{Cate} were the first authors
to combine this technique with the lattice Boltzmann method to simulate flows around moving particles. In these initial works, an empirical parameter needed to be adjusted to obtain 
satisfactory results. To circumvent this difficulty, Feng and Michaelides \cite{Feng2} and Dupuis et al. \cite{Dupuis_2008} coupled the direct forcing immersed boundary method 
developed by Fadlun et al. \cite{Fadlun_2000} in a finite difference formulation,
with the lattice Boltzmann method. However, with this method, the force that represents the influence of the immersed boundary on the fluid, is applied 
to the Eulerian 
fluid nodes surrounding the immersed boundary, and the no-slip boundary condition is not exactly satisfied at the fluid solid interface. To improve that method, 
Wu and Shu \cite{Wu_2009} developed a variant of the immersed boundary lattice Boltzmann method based on an implicit correction of the velocity calculated on 
the Eulerian
nodes in the vicinity of the immersed boundary. With that method, the no-slip boundary condition is satisfied at the fluid solid interface.\\
Meldi et al. \cite{Meldi_2013} applied the Arbitrary Lagrangian Eulerian (ALE) approach in a lattice Boltzmann framework. 
The ALE technique consists in combining the Eulerian technique far from the moving obstacle, and the Lagrangian technique in the vicinity of the 
obstacle. In the fluid region near the obstacle, the conservation equations are written on a mesh that moves arbitrarily according to 
time \cite{Hughes_1981, Donea2_2004}. This implies an increase in computing time. The Distributed Lagrange Multiplier Fictitious 
Domain (DLM/FD) method was also coupled with the LBM to calculate flows around deforming bodies \cite{Shi_2005}. This work was based on the 
DLM/FD method developed by Glowinski et al. \cite{Glowinski_1999} (for rigid body motion) and extended to deforming bodies by Yu \cite{Yu_2005}. 
The DLM/FD method consists in writing the governing equations on a fixed domain, including the fluid and the solid regions, and using Lagrange 
multipliers (or pseudo body forces) to make the fictitious fluid in the solid region move like the solid medium. With this method the costly 
re-meshing step is avoided.\\
Fluid Structure Interaction (FSI) problems can be tackled with the volume penalization technique by extending the Navier-Stokes equations 
to the whole domain (fluid domain and solid domain), and penalizing the solid domain. 
This method is efficient for fixed Cartesian grids, and it avoids the re meshing step necessary in classical approaches devoted to 
Fluid Structure Interaction modeling \cite{Souli_2001,Donea2_2004}. The promising results obtained with the Volume Penalization technique 
combined with various solving approaches (finite element method, finite volume method, spectral method, vortex method) 
\cite{Angot_1999, Maury, Morales_2014, Kadoch_2016}  make 
the volume penalization method a good candidate 
to be coupled with LBM for modeling fluid structure interaction problems. In contrast with the classical immersed boundary 
methods, the convergence of the volume penalization method has been mathematically proved \cite{Angot_1999}. To the authors' knowledge, this method has not yet been 
developed in a lattice Boltzmann framework.\\
In a previous paper \cite{Benamour_2015, Malek}, we successfully applied the combined Volume Penalization Lattice Boltzmann Method (VP-LBM)
to flows around motionless obstacles, for relatively low Reynolds numbers (less or equal than 100). The Single Relaxation Time Lattice Boltzmann Method was used. 
In this work, we combined the Multi Relaxation Time Lattice Boltzmann Method with the Volume Penalization approach, and we simulated flows for which an 
analytical solution is known. For these cases, the flow was created
by moving boundaries: an asymmetric flow and a flow between two concentric cylinders. To test the ability of the VP-LBM to calculate unsteady
flows around obstacles, we also
simulated a flow past a cylinder at a relatively high Reynolds number: $Re=500$. For that more complex case, the VP LBM method, which
is easy to implement, was parallelized
In the following paragraph, the Lattice Boltzmann method combined with the volume penalization technique will be described. Then, the numerical results obtained 
for the test cases will be presented. In the last paragraph of the paper, concluding remarks will be given.
\section{Governing equations}
\subsection{Fluid flow computation}
When solving fluid structure interaction problems using a volume penalization technique, a source term 
involving the influence of the obstacle on the fluid flow is introduced into the Navier-Stokes equations. The incompressible Navier-Stokes equations
with a source term, and the continuity equation read:
\begin{equation}
\begin{split}
\begin{array}{ll}
\displaystyle \rho \frac{\partial \boldsymbol{u}}{\partial t}+\rho \boldsymbol{u}\cdot \nabla  \boldsymbol{u}= 
- \nabla p + \mu \nabla^2 \boldsymbol{u} + \boldsymbol{f} \vspace{0.3cm} \\
\displaystyle \nabla \cdot \boldsymbol{u}=0 
\end{array}
\end{split}
\label{momentum}
\end{equation}
where $\rho$ and $\mu$ are the density and the dynamic viscosity of the fluid considered, $p$ and $\boldsymbol{u}$ are the pressure and the velocity, 
$\boldsymbol{f}$ is a body force that mimics the influence of the obstacle on the fluid flow. In this article, the lattice Boltzmann method was chosen for fluid flow 
computation. The lattice Boltzmann method consists in solving a simplified version of the Boltzmann equation on a lattice, where fluid parcels move according
to a finite set of prescribed
velocities. The correspondence between the lattice Boltzmann equation (mesoscopic level) and the Navier-Stokes equations (macroscopic level) is achieved thanks to the Chapman Enskog procedure.
\subsubsection{Single Relaxation Time LBM}
This model is based on a linear collision operator, proposed by Bhatnagar, Gross and Krook (BGK) \cite{Chen_Doolen_1998, Guo_Shu_2013}.
The lattice Boltzmann equation with a forcing term is expressed as:
\begin{equation}
 f_{\alpha}(\boldsymbol{x}+\boldsymbol{\xi}_{\alpha} \Delta t,t+ \Delta t)-f_{\alpha}(\boldsymbol{x},t)=-\frac{1}{\tau}(f_{\alpha}(\boldsymbol{x},t)-f_{\alpha}^{eq}(\boldsymbol{x},t))+\Delta t F_{\alpha}
\label{LatticeBoltz}
 \end{equation}
where $f_{\alpha}$ is the distribution function along the $\alpha$ direction, $\boldsymbol{\xi_{\alpha}}$ is the particle velocity, $f_{\alpha}^{eq}$ is the 
equilibrium distribution function, 
 $\tau$ is the non-dimensional relaxation time, $\Delta t$ is the time step, and $F_{\alpha}$ is a 
forcing term that corresponds to the source term $\boldsymbol{f}$. The nine-velocity square lattice model $D2Q9$ shown in figure 1 was chosen. In this model, the lattice velocities
are:
\begin{equation}
\begin{split}
\begin{array}{llll}
 \displaystyle \boldsymbol \xi_{\alpha}=
 \left\{
       \begin{array}{ll}
   \displaystyle (0, 0), \hspace{6.5cm}\alpha=0 \vspace{0.2cm} \\
   (cos[(\alpha-1)\pi/2], sin[(\alpha-1)\pi/2]) c, \hspace{1.55cm}\alpha=1,2,3,4 \vspace{0.2cm} \\
  (cos[(2\alpha-9)\pi/4], sin[(2\alpha-9)\pi/4]) \sqrt{2} c,  \hspace{0.65cm}\alpha=5,6,7,8 \\
 \end{array}
\right.
\end{array}
\end{split}
\label{vecteurs_D2Q9}
\end{equation} 
where $c=\Delta x/\Delta t$, and $\Delta x$ is the lattice spacing (In this paper, we choose $\Delta x=\Delta t=1$). 
%%%%%%%%%%%%%%%%%%%%%%%%%%%%%%%%%%%%%%%%%%%%%%%%%%%%%%%%%%%%%%%%%%%%%%%%%%%%%%%%%%%%%%%%%%%%%%%%%%%%%%%%%%%%%%%%%%%%%%%%%%%%%%
\begin{figure}[h!]
 \centering
 \includegraphics[width=11cm,height=6cm, angle=0]{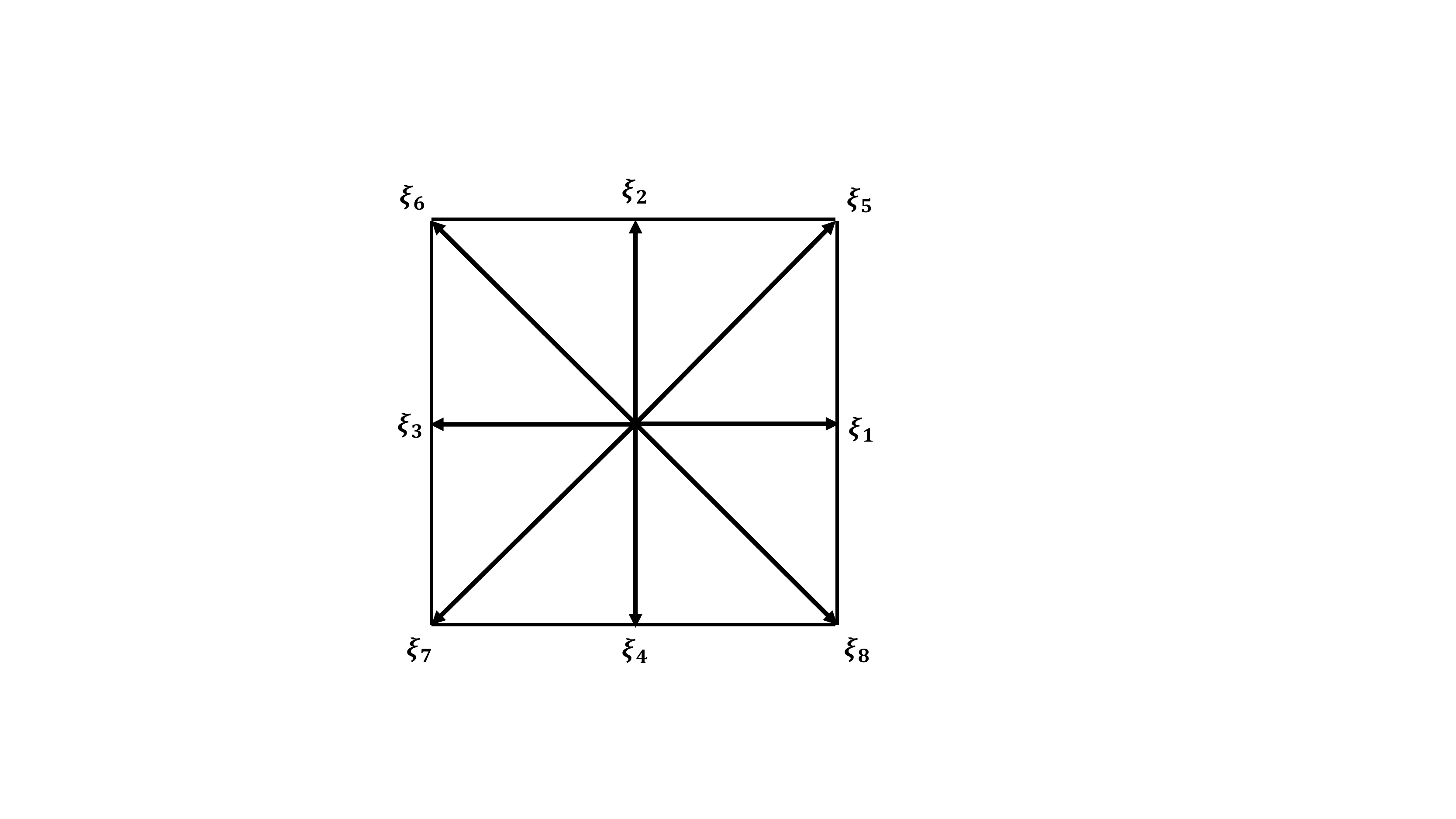}
 \caption{Discrete velocities of the $D2Q9$ model}
 \label{eq9_fig1}
 \end{figure}
%%%%%%%%%%%%%%%%%%%%%%%%%%%%%%%%%%%%%%%%%%%%%%%%%%%%%%%%%%%%%%%%%%%%%%%%%%%%%%%%%%%%%%%%%%%%%%%%%%%%%%%%%%%%%%%%%%%%%%%%%%%%%%
 For the $D2Q9$ model, the equilibrium distribution function is expressed as:
\begin{equation}
f_\alpha ^{eq} (\boldsymbol{x},t)= w_\alpha \rho \left[1+\frac{\boldsymbol{\xi}_\alpha \cdot \boldsymbol{u}}{c_s ^2}+\frac{\boldsymbol{u}\boldsymbol{u}\colon (\boldsymbol{\xi}_\alpha \boldsymbol{\xi}_\alpha-c_s ^2 I)}{2 c_s ^4}\right]
\end{equation}
where $w_{\alpha}$ is the weighting factor given by:
\begin{equation}
\begin{split}
\begin{array}{llll}
 \displaystyle w_\alpha=
 \left\{
       \begin{array}{ll}
   \displaystyle 4/9, \hspace{1.0cm}\alpha=0 \vspace{0.2cm} \\
   1/9, \hspace{1.0cm}\alpha=1,2,3,4 \vspace{0.2cm} \\
  1/36,  \hspace{0.8cm}\alpha=5,6,7,8 \\
 \end{array}
\right.
\end{array}
\end{split}
\label{anex_eq19}
\end{equation} 
$c_s$ is the speed of sound ($c_s=c/\sqrt{3}$ for the $D2Q9$ model) and $I$ is the identity matrix. The incompressible Navier-Stokes equations (\ref{momentum}) can be obtained 
from equation (\ref{LatticeBoltz}), using the Chapman-Enskog procedure. This procedure leads to the following expression
for the kinematic viscosity of the fluid:
\begin{equation}
\nu=c_s^2 \Delta t (\tau - \frac{1}{2})
\end{equation}
In the lattice Boltzmann equation, the forcing term $F_{\alpha}$, that represents the body force term $\boldsymbol{f}$ in the Navier-Stokes equations, was modeled using the
formulation proposed by Guo et al. \cite{guo:2002fp}:
\begin{equation}
 F_\alpha=\left(1-\frac{1}{2 \tau}\right) w_\alpha \left[ \frac{\boldsymbol{\xi}_\alpha -\boldsymbol{u}}{c_s ^2} +
 \frac{(\boldsymbol{\xi}_\alpha \cdot \boldsymbol{u})}{c_s ^4}\boldsymbol{\xi}_\alpha \right] \cdot \boldsymbol{f}
\label{Guo1}
\end{equation}
where the macroscopic quantities are calculated as follows:
\begin{equation}
\displaystyle \rho = \displaystyle \sum_{\alpha}f_\alpha(\boldsymbol x, t)\vspace{0.25cm}\\
\label{density}
\end{equation}
\begin{equation}
 \displaystyle \rho \boldsymbol u= \displaystyle \sum_{\alpha}\boldsymbol \xi_\alpha f_\alpha(\boldsymbol x, t)+ \frac{\Delta t}{2}\boldsymbol{f}
 \label{velocity}
\end{equation}
As demonstrated by Guo et al., this force treatment leads to the exact Navier-Stokes equations.
%%%%%%%%%%%%%%%%%%%%%%%%%%%%%%%%%%%%%%%%%%%%%%%%%%%%%%%%%%%%%%%%%%%%%%%%%%%%%%%%%%%%%%%%%%%%%%%%%%%%%%%%%%%
\subsubsection{Multiple Relaxation Time LBM}
This model, which was first introduced by d'Humières \cite{d'Humieres_1992}, is more stable than the single relaxation time lattice Boltzmann model. With this model, 
the collision process is carried out in the moment space. A transformation matrix $M$ enables to express the moments $m_{\alpha}(\boldsymbol{x},t)$ according 
to the distribution functions $f_{\alpha}(\boldsymbol{x},t)$. The lattice Boltzmann equation with a source term, as given by Lu et al. \cite{lu_immersed_2012}, becomes:
\begin{multline}
 |f_{\alpha}(\boldsymbol{x}+\boldsymbol{\xi}_{\alpha} \Delta t,t+ \Delta t)\rangle-|f_{\alpha}(\boldsymbol{x},t)\rangle=\\
 -M^{-1}[S(|m_{\alpha}(\boldsymbol{x},t)\rangle-|m_{\alpha}^{eq}(\boldsymbol{x},t)\rangle)
 -(I-S/2)M\Delta t |F_{\alpha}(\boldsymbol{x},t)\rangle]
 \label{equa_lbl_mrt}
\end{multline}
In this equation, $|\bullet \rangle$ denotes a column vector and $I$ is the identity matrix. The moments $|m_{\alpha}(\boldsymbol{x},t)\rangle=(m_0,m_1,\ldots,m_8)^T$ are deduced from:
\begin{equation}
 |m_{\alpha}(\boldsymbol{x},t)\rangle=M|f_{\alpha}(\boldsymbol{x},t)\rangle \Rightarrow |f_{\alpha}(\boldsymbol{x},t)\rangle=M^{-1}|m_{\alpha}(\boldsymbol{x},t)\rangle
\end{equation}
The corresponding equilibria of the moments are given by:
\begin{equation}
 |m_{\alpha}^{eq}(\boldsymbol{x},t)\rangle=\rho(1,-2+3(u^2+v^2),1-3(u^2+v^2),u,-u,v,-v,u^2-v^2,uv)^T
\end{equation}
The transformation matrix $M$ is:
\begin{equation}
\begin{split}
\begin{array}{cccccccccccccc}
 \displaystyle M=
\begin{pmatrix}
1 & \hspace{0.2cm} 1 \hspace{0.2cm} & 1 \hspace{0.2cm} & 1 \hspace{0.2cm} & \hspace{0.2cm} 1 & \hspace{0.2cm} 1 & \hspace{0.2cm} 1 & \hspace{0.2cm} 1 &\hspace{0.2cm} 1   \\
-4 & -1 & -1 & -1 & -1 & 2 & 2 & 2 & 2 \\
4 & -2 & -2 & -2 & -2 & 1 & 1 & 1 & 1 \\
0 & 1 & 0 & -1 & 0 & 1 & -1 & -1 & 1 \\
0 & -2 & 0 & 2 & 0 & 1 & -1 & -1 & 1 \\
0 & 0 & 1 & 0 & -1 & 1 & 1 & -1 & -1 \\
0 & 0 & -2 & 0 & 2 & 1 & 1 & -1 & -1 \\
0 & 1 & -1 & 1 & -1 & 0 & 0 & 0 & 0 \\
0 & 0 & 0 & 0 & 0 & 1 & -1 & 1 & -1 
\end{pmatrix}
 \end{array}
\end{split}
\label{eq1.7}
\end{equation}
$S$ is a diagonal
 matrix that contains the relaxation rates of each moment:\\
\begin{equation}
 S=diag (s_0,s_1,s_2,s_3,s_4,s_5,s_6,s_7,s_8)=diag(0,s_n,s_n,0,s_q,0,s_q,s_n,s_n)
\end{equation}
Ginzburg and d'Humières \cite{Ginzburg_2003} showed that, for straight boundaries, the numerical slip at the fluid-solid interface can be avoided if the following relationship is satisfied:
\begin{equation}
 \Lambda=\left(\frac{1}{s_n}-\frac{1}{2}\right)\left(\frac{1}{s_q}-\frac{1}{2}\right)
 \label{lambda_parameter}
\end{equation}
with $\Lambda=\displaystyle\frac{3}{16}$ and $s_n=\displaystyle\frac{1}{\tau}$.
The last term in equation (\ref{equa_lbl_mrt})
represents the forcing term developed by Lu et al. \cite{lu_immersed_2012} for the MRT LBM, which is a general formulation of the forcing term proposed 
by Guo et al. \cite{guo:2002fp} for the SRT LBM. In this expression, $F_{\alpha}$ is calculated according to:
\begin{equation}
 F_{\alpha}=w_{\alpha}\left(\frac{\boldsymbol{\xi}_{\alpha}-\boldsymbol{u}}{c_s^2}+
 \frac{(\boldsymbol{\xi}_{\alpha}\cdot\boldsymbol{u})}{c_s^4}\boldsymbol{\xi}_{\alpha}\right)\cdot \boldsymbol{f}
 \label{Guo2}
\end{equation}
When all relaxation rates are equal to $\displaystyle\frac{1}{\tau}$, the MRT model leads to the SRT model, including the forcing term developed 
by Guo et al. \cite{guo:2002fp}.
%%%%%%%%%%%%%%%%%%%%%%%%%%%%%%%%%%%%%%%%%%%%%%%%%%%%%%%%%%%%%%%%%%%%%%%%%%%%%%%%%%%%%%%%%%%%%%%%%%
\subsection{Volume penalization method}
\label{sec:VP}
The aim of this work is to apply the volume penalization method in a lattice Boltzmann framework to take into account the presence of an obstacle in a flow.
In the volume penalization method, the body force
term in the Navier-Stokes equations (\ref{momentum}) applies to all Cartesian nodes of the whole computational domain (fluid domain and solid domain). The volume 
source term reads:
\begin{equation}
 \boldsymbol{f}=-\rho\frac{\chi}{\eta}(\boldsymbol{u}-\boldsymbol{u}_s)
\end{equation}
where $\chi$ is a mask function equal to $0$ in the fluid domain, and $1$ in the solid domain, $\eta$ is a permeability or penalization parameter, that is very small (for all
computations, a value of $\eta=10^{-7}$ was chosen), and 
$\boldsymbol{u}_s$ is the solid velocity. With this method, the penalization term is very high in 
the solid domain, hence the velocity is equal to the solid velocity, and it is null in the fluid domain.
This penalization term is introduced into the lattice Boltzmann equation (\ref{LatticeBoltz}) (SRT lattice Boltzmann equation) or (\ref{equa_lbl_mrt}) (MRT lattice Boltzmann
equation), using the forcing term proposed by Guo et al. (\ref{Guo1}) (for the SRT LBM)
 or (\ref{Guo2}) (for the MRT LBM). The lattice Boltzmann equation is solved on both fluid and solid domains. 
Equation (\ref{velocity}) is used to calculate the velocity in
the fluid and solid domains. The right hand side of this equation contains the fluid velocity via the penalization term $\boldsymbol{f}$. This leads to the following expression
for the velocity:
\begin{equation}
\begin{split}
\begin{array}{ll}
\displaystyle \boldsymbol{u} = \displaystyle \frac{ \displaystyle \sum_{\alpha}\boldsymbol{\displaystyle \xi_{\alpha}}f_{\alpha} + \displaystyle \frac{\Delta t }{2} \frac{\displaystyle \chi}{\eta} \rho\boldsymbol{u_{s}}}{\displaystyle \displaystyle \rho+ \frac{\Delta t }{2} \frac{\displaystyle \chi}{\displaystyle \eta}\rho }
\end{array}
\end{split}
\label{vitesse_fluide_penalisation_1}
\end{equation}
Since the velocity boundary condition is naturally applied at the solid obstacle thanks to the mask function, this method is easy to implement.
%%%%%%%%%%%%%%%%%%%%%%%%%%%%%%%%%%%%%%%%%%%%%%%%%%%%%%%%%%%%%%%%%%%%%%%%%%%%%%%%%%%%%%%%%%%%%%%%%%%%%%%%%%%%%
%%%%%%%%%%%%%%%%%%%%%%%%%%%%%%%%%%%%%%%%%%%%%%%%%%%%%%%%%%%%%%%%%%%%%%%%%%%%%%%%%%%%%%%%%%%%%%%%%%%%%%%%%%%%%
\section{Numerical results and discussion}
In this section, we present numerical results obtained with the VP-LBM. To validate this method, we selected flows for which an analytical solution
exists: a symmetric shear flow between two moving plates, and a Couette flow between two cylinders. These 
test cases were chosen by Le and Zhang \cite{Le_Zhang_2009} and by Lu et al. \cite{lu_immersed_2012} for evaluating
the immersed boundary approach combined with the lattice Boltzmann method. Our results will therefore also be compared with the results computed by Lu et al. 
with the IB-LBM (Immersed Boundary Lattice Boltzmann Method). Since for these configurations the Reynolds numbers were low, the VP-LBM was also
applied to the computation of a bluff body flow at a higher Reynolds number (e.g. 500), and it will be showed that satisfactory results were obtained.
%%%%%%%%%%%%%%%%%%%%%%%%%%%%%%%%%%%%%%%%%%%%%%%%%%%%%%%%%%%%%%%%%%%%%%%%%%%%%%%%%%%%%%%%%%%%%%%%%%%%%%%%%%%%%
%%%%%%%%%%%%%%%%%%%%%%%%%%%%%%%%%%%%%%%%%%%%%%%%%%%%%%%%%%%%%%%%%%%%%%%%%%%%%%%%%%%%%%%%%%%%%%%%
\subsection{Symmetric shear flow}
First, we focused on a symmetric shear flow, between two moving parallel plates (see figure 2) with a thickness $e=50 \;l.u.$
%%%%%%%%%%%%%%%%%%%%%%%%%%%%%%%%%%%%%%%%%%%%%%%%%%%%%%%%%%%%%%%%%%%%%%%%%%%%%%%%%%%%%%%%%%%%%%%%%%%%%%%%%%%%%%%%%%%%%%%%%%%%%%%%%%%%%%%%%%%%%%%%%%%%%%%
\begin{figure}[h!]
 \centering
 \includegraphics[width=11cm,height=5.5cm, angle=0]{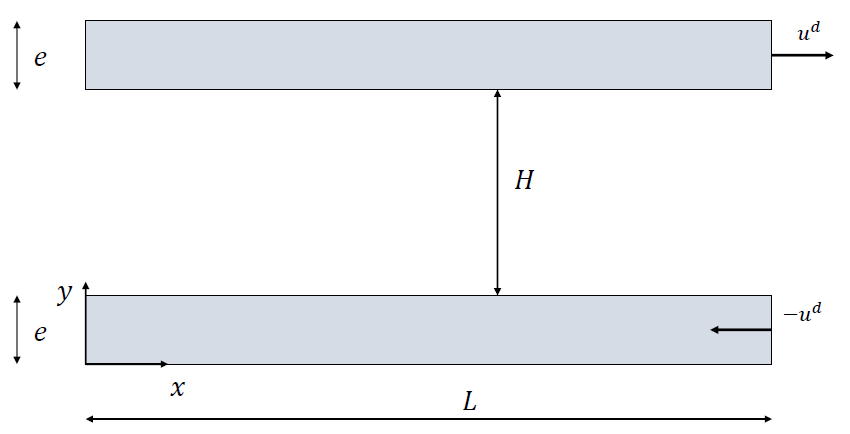}
 \caption{The symmetric shear flow case}
 \label{Couet_fig1}
 \end{figure}
%%%%%%%%%%%%%%%%%%%%%%%%%%%%%%%%%%%%%%%%%%%%%%%%%%%%%%%%%%%%%%%%%%%%%%%%%%%%%%%%%%%%%%%%%%%%%%%%%%%%%%%%%%%%%%%%%%%%%%%%%%%%%%%%%%%%%%%%%%%%%%%%%%%%%%%
This configuration was modeled by Le and Zhang \cite{Le_Zhang_2009}, Lu et al. \cite{lu_immersed_2012}, and Farnoush and Manzari \cite{Farnoush_2014},
who analyzed the immersed boundary method combined with the lattice Boltzmann method. The length of the domain is $L=200 \;l.u.$, and the distance between the two plates is $H=100 \;l.u.$ The upper and lower plates move according to $x$ direction,
with velocities $u^d$ and $-u^d$, respectively. Periodic boundary conditions were applied according to $x$ and $y$ directions. 
For this configuration, the lattice Boltzmann simulations were performed with a maximum lattice velocity that ensured that the Mach number was small ($u^d=0.01 \;l.u.$). 
For this steady unidirectional flow in the absence of a pressure gradient, the analytical solution of the Navier-Stokes 
equations is:
\begin{equation}
\begin{split}
\begin{array}{llll}
\displaystyle u_{analytical}(y) = \frac{2 u^d}{H}(y-e) - u^d   
\end{array}
\end{split}
\label{analy}
\end{equation}
%%%%%%%%%%%%%%%%%%%%%%%%%%%%%%%%%%%%%%%%%%%%%%%%%%%%%%%%%%%%%%%%%%%%%%%%%%%%%%%%%%%%%%%%%%%%%%%%%%%%%%%%%%%%%%%%%%%%%%%%%%%%%%%%%%%%%%%%%%%%%%%%%%%%%%%
\begin{figure}[h!]
  \begin{center}
    \leavevmode 
    \subfloat[influence of $\tau$ (VP SRT LBM)]{
      \includegraphics[width=6.5cm, height=5cm]{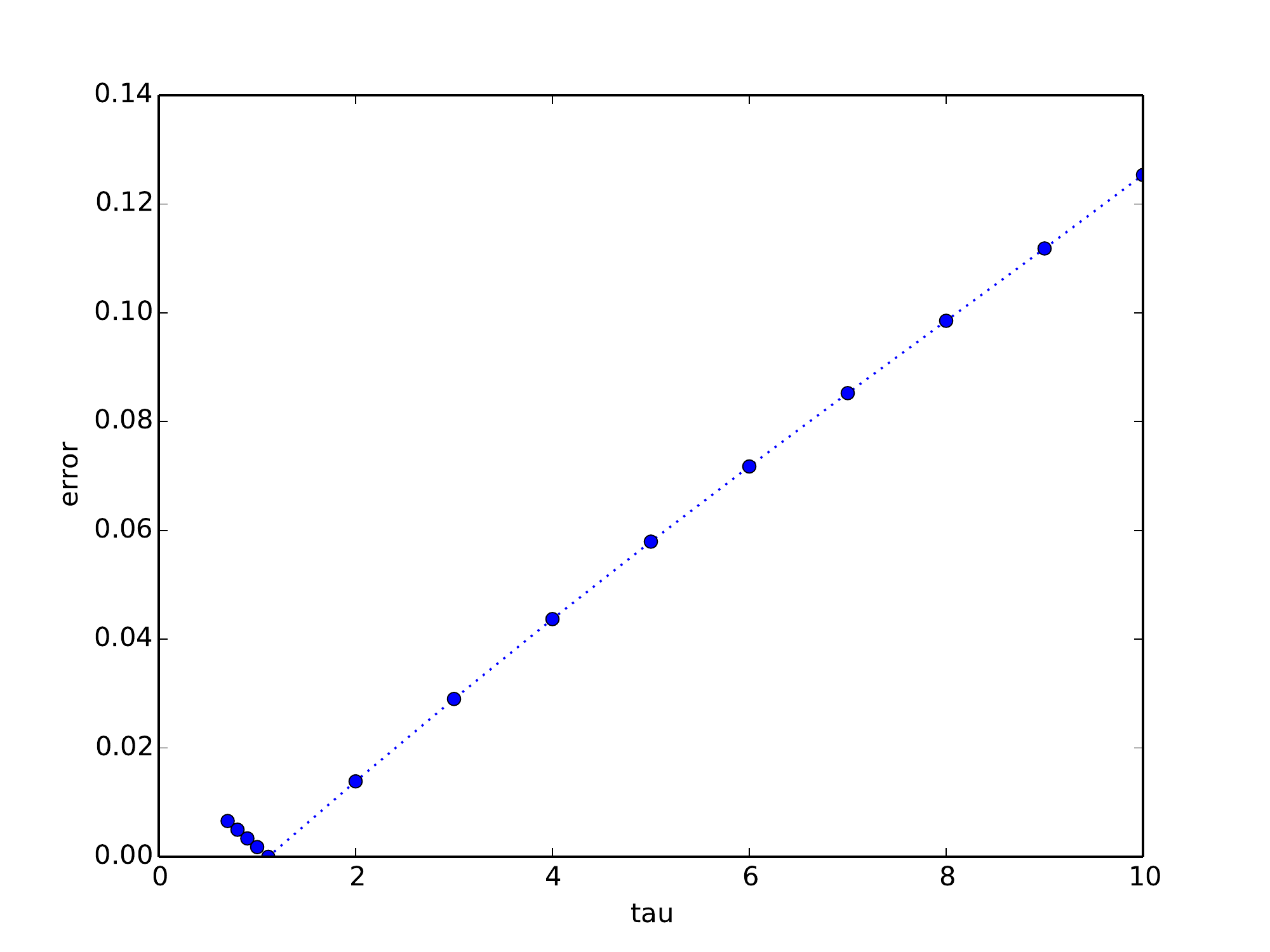}}     
      \hspace{-0.5cm}   
    \subfloat[influence of $\Lambda$ (VP MRT LBM)]{
      \includegraphics[width=6.5cm, height=5cm]{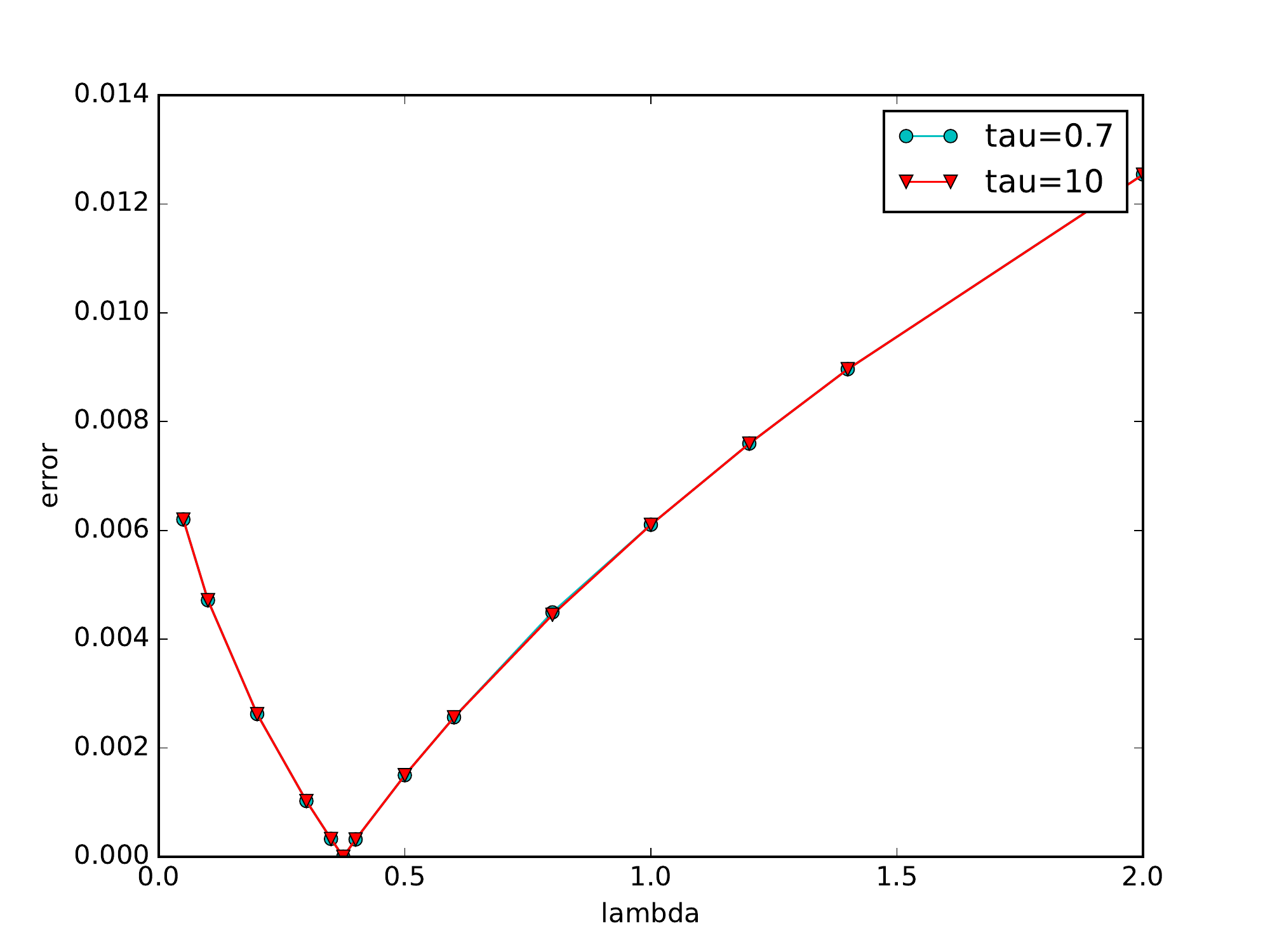}}        
       \vspace{-0.065cm}            
       \caption{(a) Influence of $\tau$ on the relative $L_2$ error of $u$, for the symmetric shear flow (VP SRT LBM); (b) Influence of $\Lambda$ 
       on the relative $L_2$ error of $u$, for the symmetric shear flow (VP MRT LBM).}
    \label{error_L2_sym}
  \end{center}
\end{figure}
%%%%%%%%%%%%%%%%%%%%%%%%%%%%%%%%%%%%%%%%%%%%%%%%%%%%%%%%%%%%%%%%%%%%%%%%%%%%%%%%%%%%%%%%%%%%%%%%%%%%%%%%%%%%%%%%%%%%%%%%%%%%%%%%%%%%%%%%%%%%%%%%%%%%%%%%
The relative $L_2$ error of $u$ in the fluid domain was calculated according to :
\begin{equation}
\begin{split}
\begin{array}{ll}
\displaystyle error = \frac{\displaystyle\sqrt{\displaystyle \sum  \left(u-u_{analytical}\right) ^2}}{\displaystyle\sqrt{\displaystyle \sum(u_{analytical})^2}}
\end{array}
\end{split}
\label{err}
\end{equation}
where the summation was performed in the fluid domain, and $u$ and $u_{analytical}$ are respectively the computed horizontal velocity and the analytical solution. 
Figure 3 compares the relative $L_2$ error of $u$ in the fluid domain, computed when using the volume penalization method and the single 
relaxation time lattice Boltzmann method (VP SRT) on 
the one hand, and the multiple relaxation time lattice Boltzmann method (VP MRT) on the other hand. In this figure, we can see that the error was very close to 
zero for 
$\tau=\frac{1}{2}+\frac{\sqrt{6}}{4}\approx 1.11$ when the flow was computed with the single relaxation time LBM, and for $\Lambda=\frac{3}{8}=0.375$ when 
the multiple relaxation
time LBM was applied.\\
%%%%%%%%%%%%%%%%%%%%%%%%%%%%%%%%%%%%%%%%%%%%%%%%%%%%%%%%%%%%%%%%%%%%%%%%%%%%%%%%%%%%%%%%%%%%%%%%%%%%%%%%%%%%%%%%%%%%%%%%%%%%%%%%%%%%%%%%%%%%%%%%%%%%%%%%%%%
\begin{figure}[h!]
  \begin{center}
    \leavevmode 
    \subfloat[difference between the numerical (VP SRT LBM) and the exact horizontal velocities]{
      \includegraphics[width=6.5cm, height=5cm]{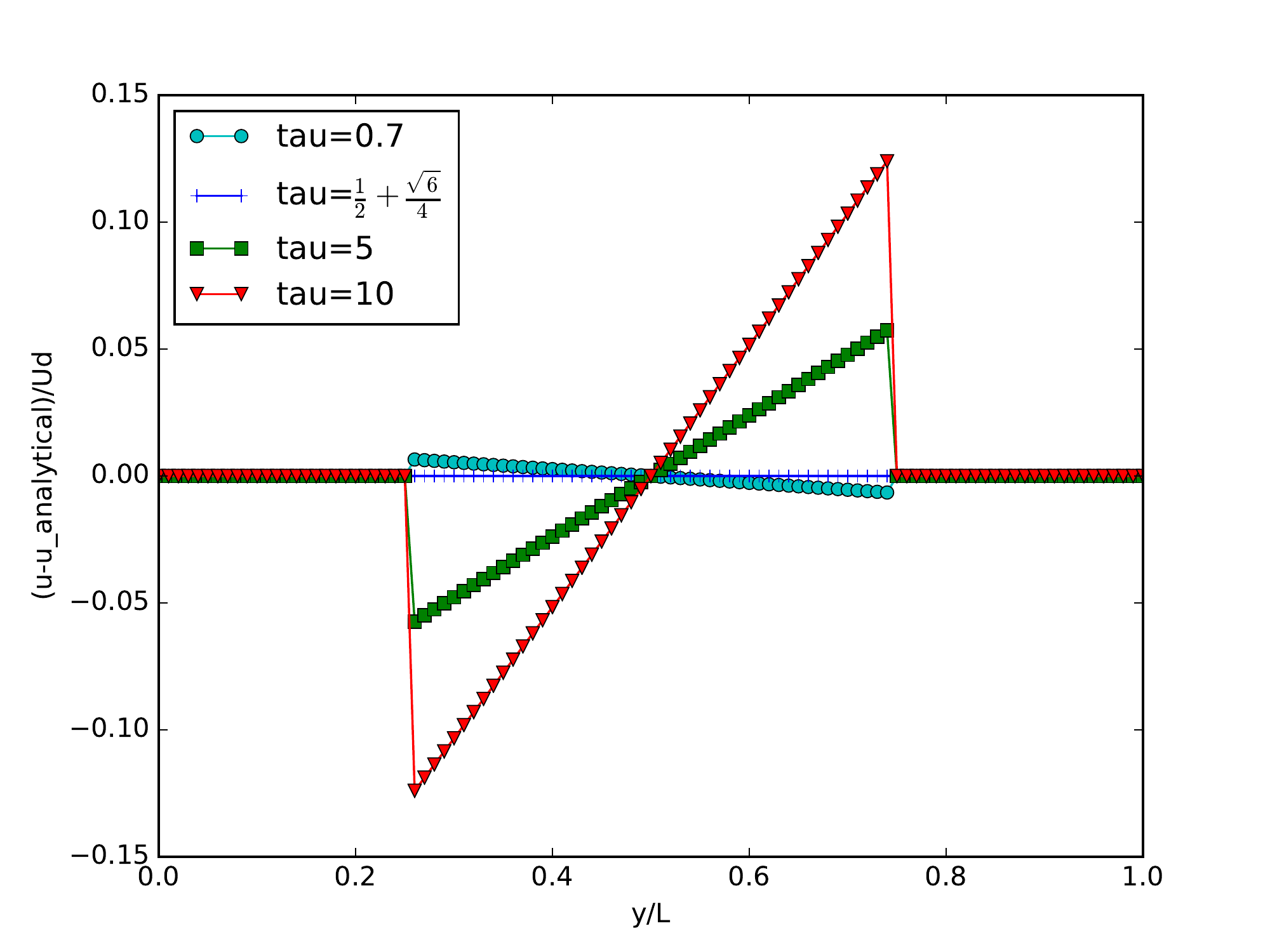}}     
      \hspace{0.5cm}   
    \subfloat[horizontal velocity obtained with the VP SRT LBM]{
      \includegraphics[width=6.5cm, height=5cm]{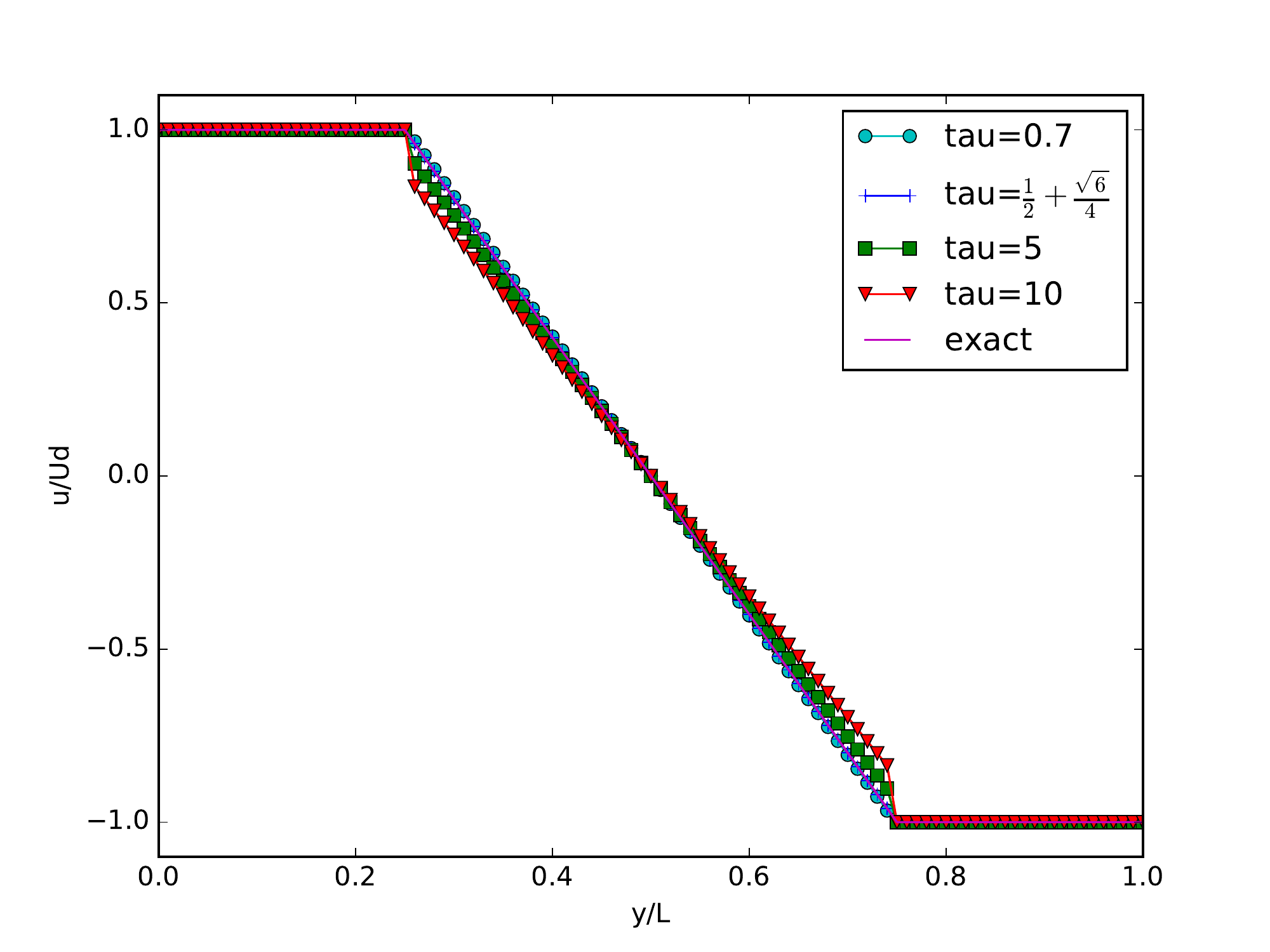}}        
       \vspace{-0.065cm}
    \subfloat[difference between the numerical (VP MRT LBM) and the exact horizontal velocities]{
       \includegraphics[width=6.5cm, height=5cm]{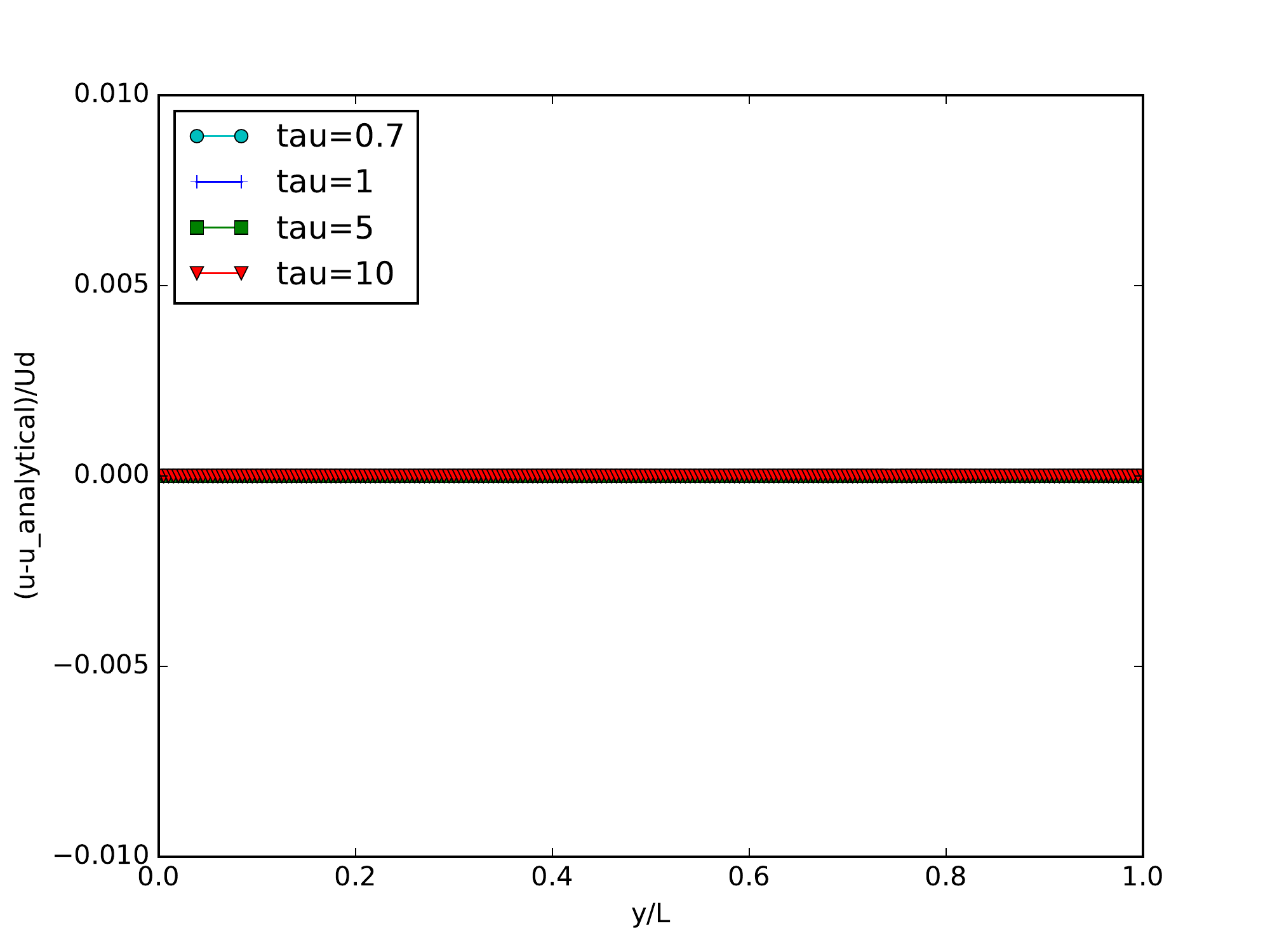}}      
       \hspace{0.5cm}   
    \subfloat[horizontal velocity obtained with the VP MRT LBM]{
       \includegraphics[width=6.5cm, height=5cm]{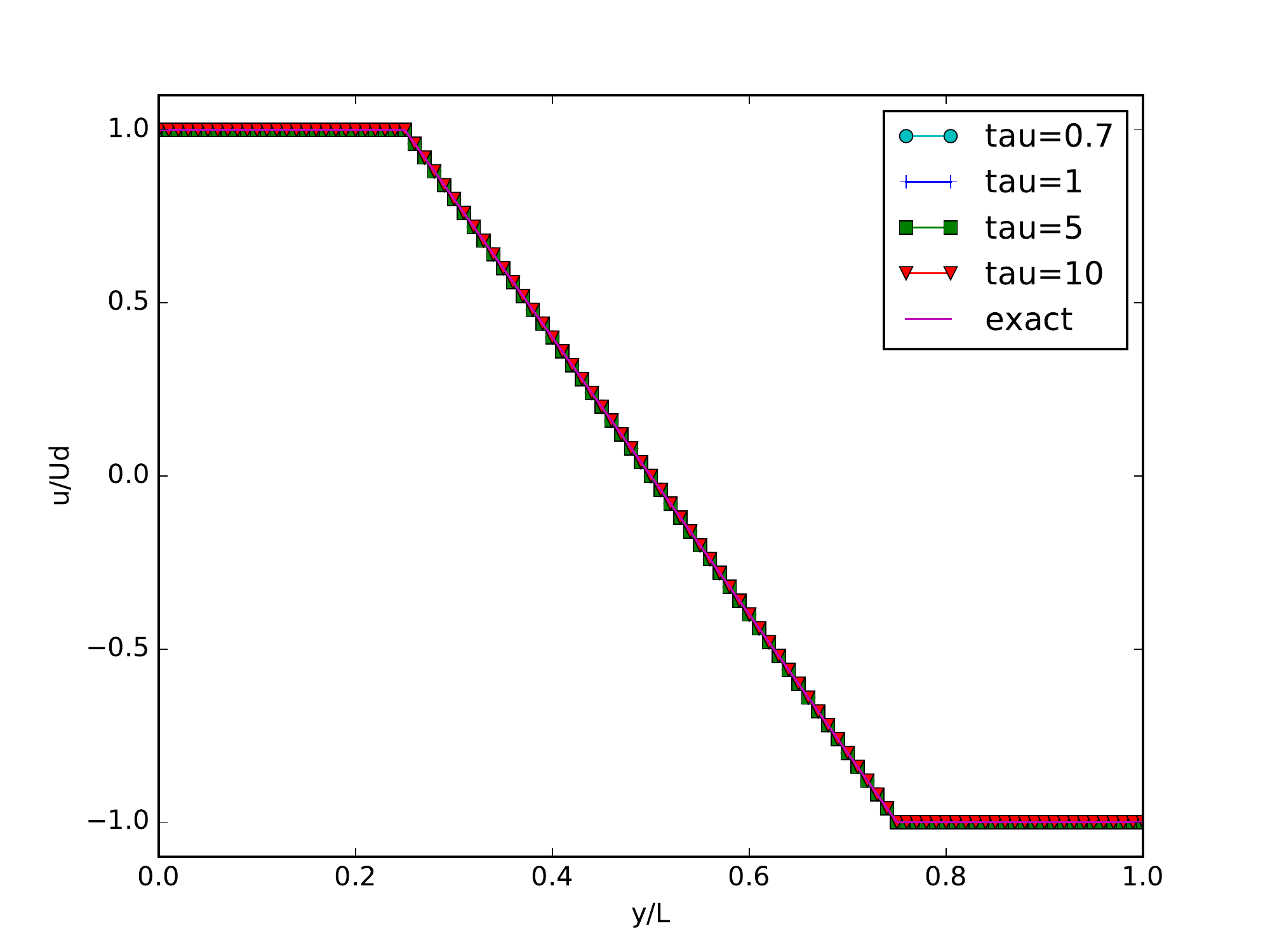}}            
       \caption{Difference between the numerical and the exact velocities (a) and (c), and velocity profiles (b) and (d), obtained at $x = L/2$, for the symmetric shear flow. For 
       the VP MRT LBM computations, $\Lambda=\frac{3}{8}$ was used. }
    \label{Sym}
  \end{center}
\end{figure}
%%%%%%%%%%%%%%%%%%%%%%%%%%%%%%%%%%%%%%%%%%%%%%%%%%%%%%%%%%%%%%%%%%%%%%%%%%%%%%%%%%%%%%%%%%%%%%%%%%%%%%%%%%%%%%%%%%%%%%%%%%%%%%%%%%%%%%%%%%%%%%%%%%
\begin{figure}[h!]
	\centering
			\includegraphics[width=6.5cm, height=5cm]{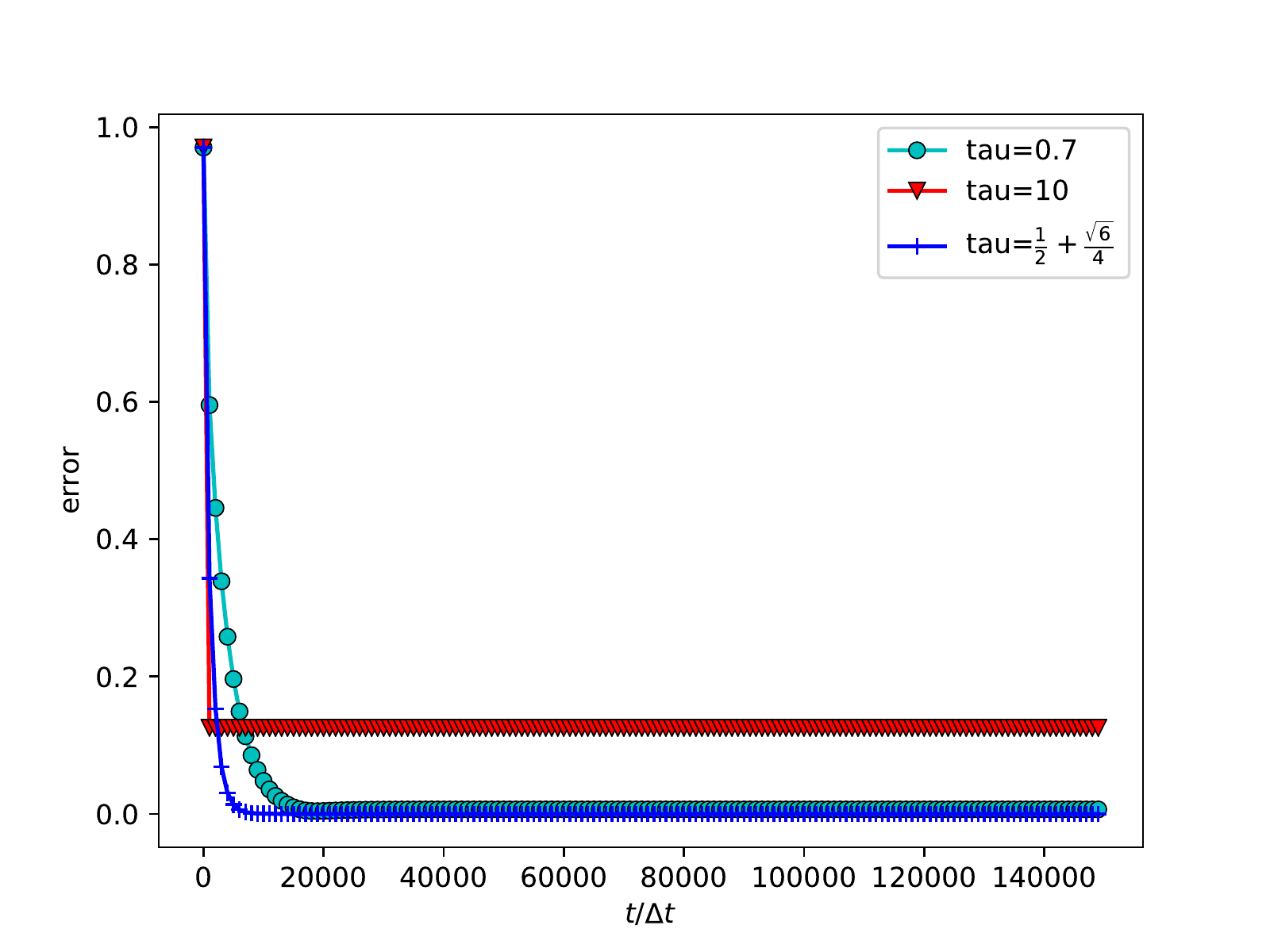}     
		\caption{Convergence history of the VP SRT LBM, for the symmetric shear flow case.}
		\label{histoire_globale_symetrique}
\end{figure}
%%%%%%%%%%%%%%%%%%%%%%%%%%%%%%%%%%%%%%%%%%%%%%%%%%%%%%%%%%%%%%%%%%%%%%%%%%%%%%%%%%%%%%%%%%%%%%%%%%%%%%%%%%%%%%%%%%%%%%%%%%%%%%%%%%%%%%%%%%%%%%%%%%
\begin{figure}[h!]
	\begin{center}
		\leavevmode 
		\subfloat[$\tau=0.7$]{
			\includegraphics[width=6.5cm, height=5cm]{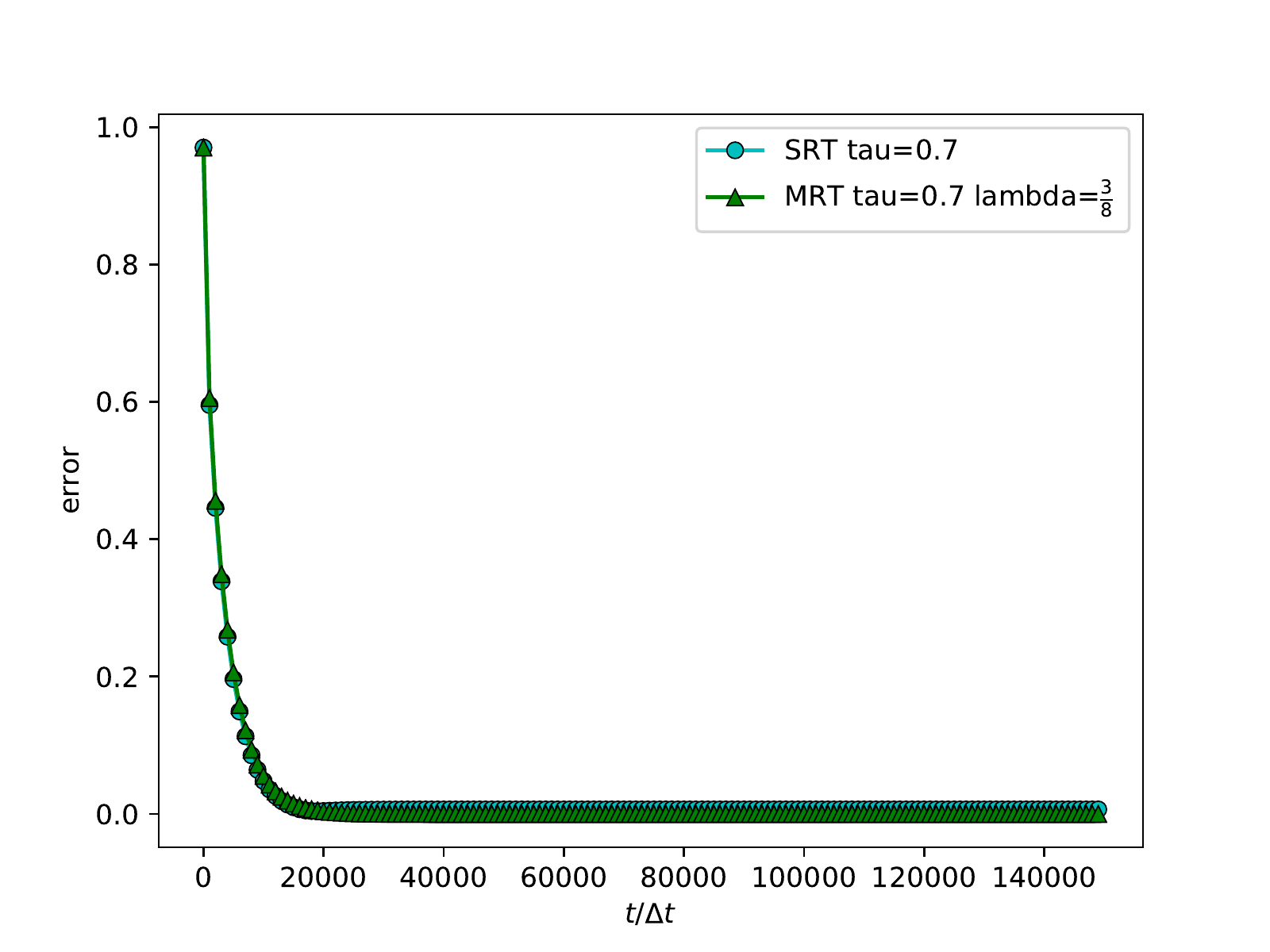}}     
		\hspace{0.5cm}   
		\subfloat[$\tau=10$]{
			\includegraphics[width=6.5cm, height=5cm]{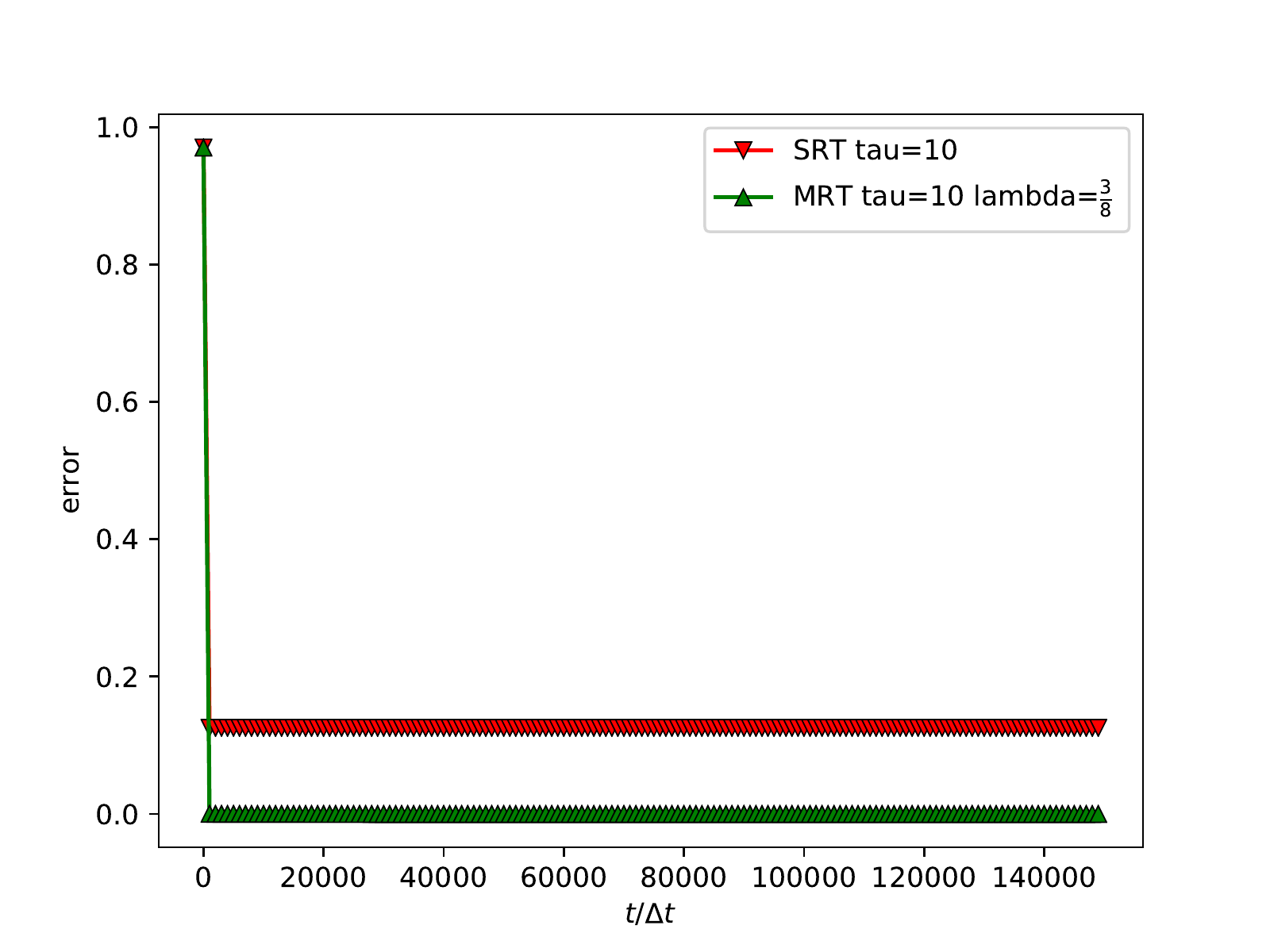}}        
		\caption{Convergence history of the VP MRT LBM for the symmetric shear flow case ($\tau=0.7$ and $\tau=10$).}
		\label{histoire_symetrique_MRT07}
	\end{center}
\end{figure}
%%%%%%%%%%%%%%%%%%%%%%%%%%%%%%%%%%%%%%%%%%%%%%%%%%%%%%%%%%%%%%%%%%%%%%%%%%%%%%%%%%%%%%%%%%%%%%%%%%%%%%%%%%%%%%%%%%%%%%%%%%%%%%%%%%%%%%%%%%%%%%%%%%
\begin{figure}[h!]
	\centering
			\includegraphics[width=6.5cm, height=5cm]{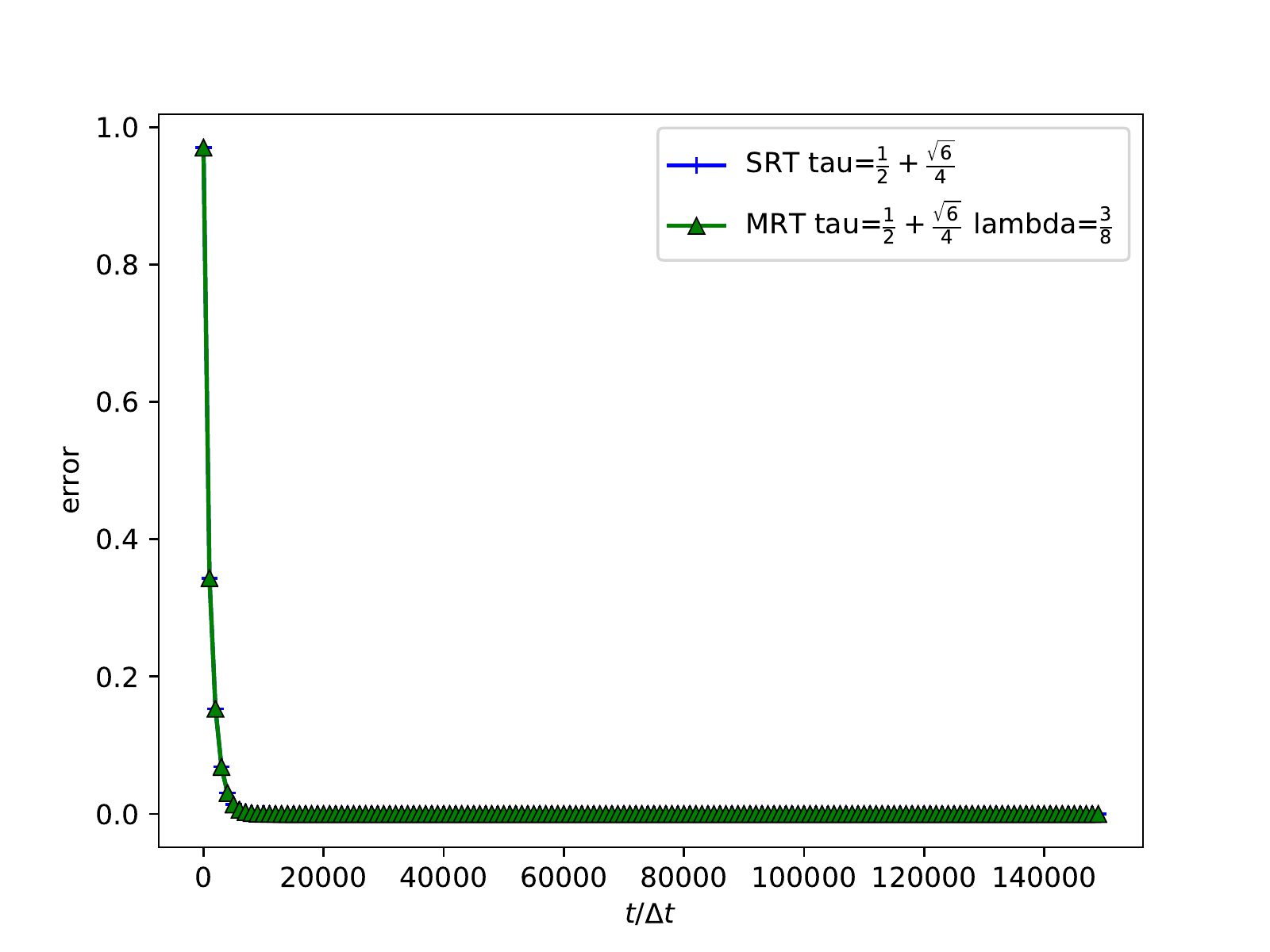}     
		\caption{Convergence histories of the VP SRT LBM ($\tau=\frac{1}{2}+\frac{\sqrt{6}}{4}$), and of the 
		VP MRT LBM ($\tau=\frac{1}{2}+\frac{\sqrt{6}}{4}$, $\Lambda=\frac{3}{8}$), for the symmetric shear flow.}
		\label{histoire_symetrique_MRT}
\end{figure}
%%%%%%%%%%%%%%%%%%%%%%%%%%%%%%%%%%%%%%%%%%%%%%%%%%%%%%%%%%%%%%%%%%%%%%%%%%%%%%%%%%%%%%%%%%%%%%%%%%%%%%%%%%%%%%%%%%%%%%%%%%%%%%%%%%%%%%%%%%%%%%%%%%%%%%%%%%%
\begin{figure}[h!]
	\begin{center}
		\leavevmode 
		\subfloat[SRT LBM]{
			\includegraphics[width=6.5cm, height=5cm]{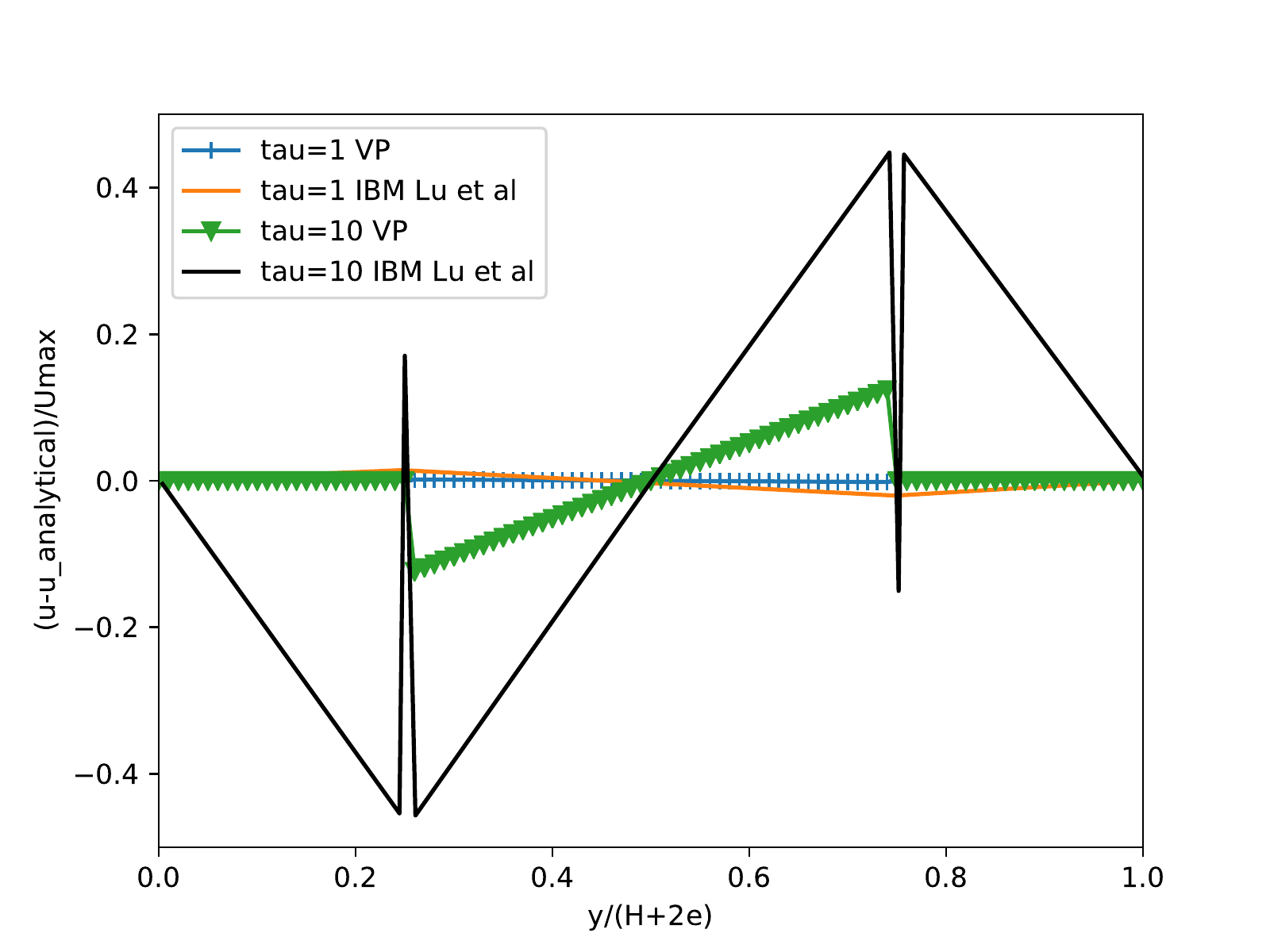}}     
		\hspace{0.5cm}   
		\subfloat[MRT LBM]{
			\includegraphics[width=6.5cm, height=5cm]{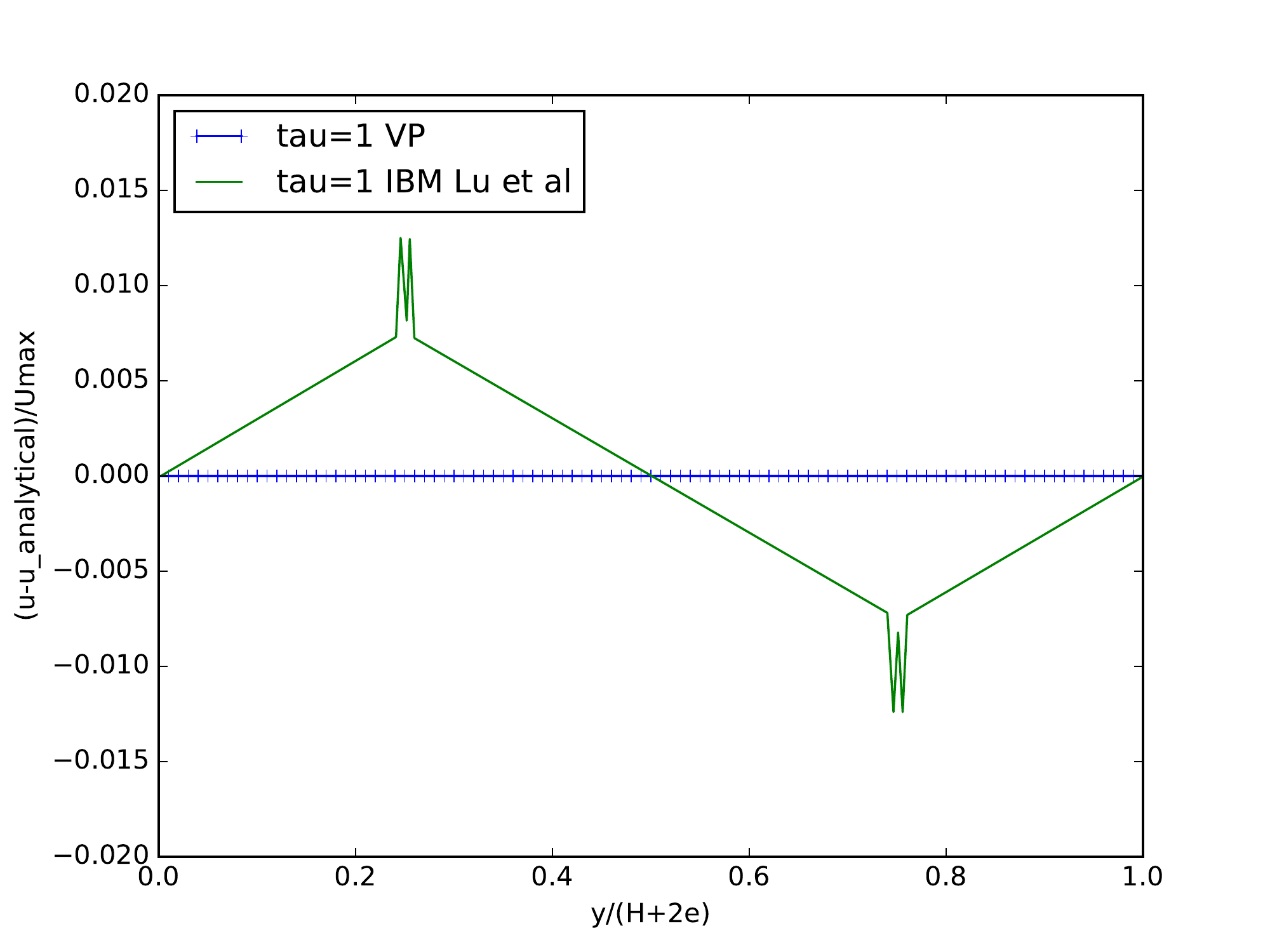}}        
		\caption{Comparison of the errors obtained with the volume penalization LBM, with those obtained with an immersed boundary
		LBM by Lu et al. \cite{lu_immersed_2012}, for the symmetric shear flow case.}
		\label{comparaison_guo_sym}
	\end{center}
\end{figure}
%%%%%%%%%%%%%%%%%%%%%%%%%%%%%%%%%%%%%%%%%%%%%%%%%%%%%%%%%%%%%%%%%%%%%%%%%%%%%%%%%%%%%%%%%%%%%%%%%%%%%%%%%%%%%%%%%%%%%%%%%%%%%%%%%%%%%%%%%%%%%%%%%%%%%%%%%%
In figure 4, the difference between the exact solution and the numerical results obtained with the SRT LBM (figure (a)) and the MRT LBM (figure (c)), and 
the corresponding velocity profiles at $x=L/2$ (figure (b) for the SRT LBM and figure (d) for the MRT LBM with $\Lambda=\frac{3}{8}$) are presented. We can see that the velocity calculated in the 
solid domain is the solid velocity naturally applied thanks to the penalization technique. For the Multiple Relaxation Time LBM with $\Lambda=\frac{3}{8}$, the solution is very close to the exact solution
for all relaxation times.\\
Figures 5, 6, and 7 show that converged results 
with an error very close to zero were obtained with the smallest number of
iterations when using the VP SRT LBM with $\tau=\frac{1}{2}+\frac{\sqrt{6}}{4}$ or the VP MRT LBM with $\tau=\frac{1}{2}+\frac{\sqrt{6}}{4}$ 
and $\Lambda=\frac{3}{8}$. In figure 7, we can also 
remark that using the VP SRT LBM with $\tau=\frac{1}{2}+\frac{\sqrt{6}}{4}$ or
the VP MRT LBM with $\Lambda=\frac{3}{8}$ leads to the same error, which is very close to zero.  Indeed, for the SRT LBM, 
we have: $s_n=s_q=\frac{1}{\tau}$, this implies: $\Lambda=(\tau-\frac{1}{2})^2=\frac{3}{8}$, 
and thus: $\tau=\frac{1}{2}+\frac{\sqrt{6}}{4}$. Furthermore, as mentioned previously, Ginzburg and d'Humières \cite{Ginzburg_2003} showed that, 
when applying the half way bounce 
back boundary condition, exact solutions can be obtained
for Poiseuille and Couette flows, if $\Lambda=\frac{3}{16}$. Lu et al. \cite{lu_immersed_2012}, who conducted a theoretical analysis on the 
Immersed Boundary Lattice Boltzmann Method, showed that, for similar flow cases, the error can be decreased but not eliminated for $\Lambda=\frac{9}{8}$.\\
For this case, we also compared the results given by the volume penalization method, with those obtained with the immersed boundary method by Lu et al. \cite{lu_immersed_2012} on
 the same grid (see figure 8). In this case involving straight boundaries, it can be noticed that for 
high relaxation times, the immersed boundary method combined with the single relaxation time LBM results in a higher error as compared with the results obtained
with the volume penalization method (figure 8(a)). We can also remark that while the error is very close to zero when using the volume penalization method together with the multiple
relaxation time LBM, it cannot be eliminated when implementing the immersed boundary method in a multiple relaxation time LBM (figure 8(b)).
%%%%%%%%%%%%%%%%%%%%%%%%%%%%%%%%%%%%%%%%%%%%%%%%%%%%%%%%%%%%%%%%%%%%%%%%%%%%%%%%%%%%%%%%%%%%%%%%%%%%
\subsection{Flow between two cylinders}
%%%%%%%%%%%%%%%%%%%%%%%%%%%%%%%%%%%%%%%%%%%%%%%%%%%%%%%%%%%%%%%%%%%%%%%%%%%%%%%%%%%%%%%%%%%%%%%%%%%%%%%%%%%%%%%%%%%%%%%%%%%%%%%%%%%%%%%%%%%%%%%%%%%%%%%%%%%%
\begin{figure}[h!]
	\centering
	\includegraphics[width=9cm,height=8cm, angle=0]{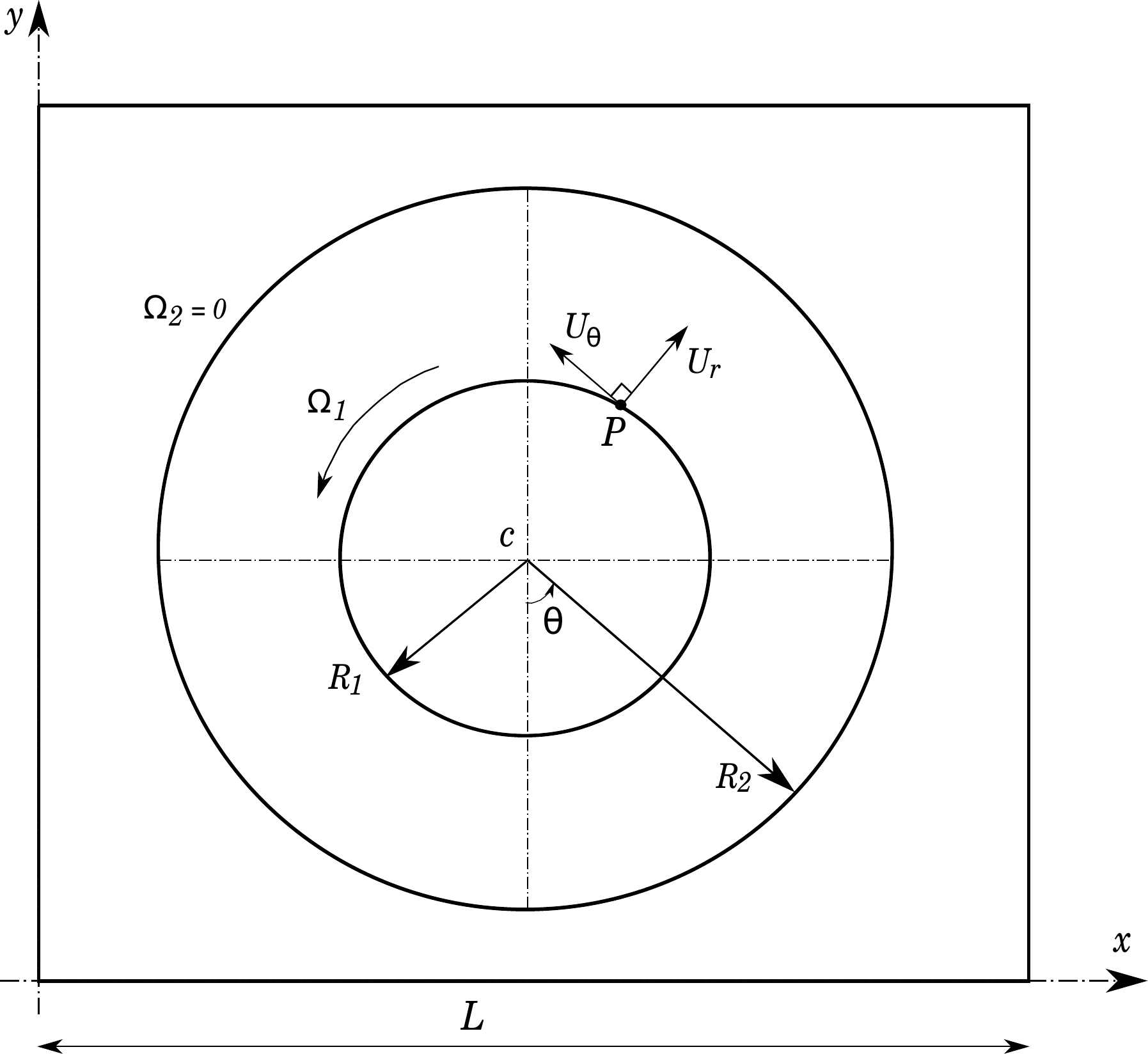}
	\caption{The cylindrical Couette flow case.}
	\label{Cylindre1}
\end{figure}
%%%%%%%%%%%%%%%%%%%%%%%%%%%%%%%%%%%%%%%%%%%%%%%%%%%%%%%%%%%%%%%%%%%%%%%%%%%%%%%%%%%%%%%%%%%%%%%%%%%%%%%%%%%%%%%%%%%%%%%%%%%%%%%%%%%%%%%%%%%%%%%%%%%%%%%%%%%%%
In this part of the study, we applied the VP LBM to a curved boundary, by focusing on a flow between two cylinders, as shown in Figure 9. 
The length and width of the square computational domain are $L=200 \;l.u.$, and the radii of the inner and the outer cylinders located in the 
center of the domain, are $R_1=45 \;l.u.$ and $R_2=70 \;l.u.$ respectively. While the outer cylinder is motionless ($\Omega_2=0$), the inner 
cylinder moves according to the angular 
velocity $\Omega_1=u^d/R_1$. In the computations, $u^d$ was small enough to deal with a low Mach number flow ($u^d=0.01 \;l.u.$). This configuration was 
studied by \cite{Le_Zhang_2009} and \cite{lu_immersed_2012}, who evaluated the immersed boundary method implemented in a lattice Boltzmann framework. 
For this case, the analytical profile of the tangential velocity component is:
\begin{equation}
\displaystyle U_{\theta \; analytical} = \frac{\displaystyle -u^d R_1} {\displaystyle (R_2^2 -R_1^2)} \hspace{0.1cm} r +\frac{\displaystyle u^d R_1 R_2^2}{\displaystyle (R_2^2 -R_1^2)} \hspace{0.1cm} \displaystyle \frac{1}{r}
\label{Sltan}
\end{equation}
where $r$ is the cylindrical coordinate.\\
%%%%%%%%%%%%%%%%%%%%%%%%%%%%%%%%%%%%%%%%%%%%%%%%%%%%%%%%%%%%%%%%%%%%%%%%%%%%%%%%%%%%%%%%%%%%%%%%%%%%%%%%%%%%%%%%%%%%%%%%%%%%%%%%%%%%%%%%%%%%%%%%%%%%%%%%%%%%%
\begin{figure}[h!]
	\begin{center}
		\leavevmode 
		\subfloat[VP SRT LBM ($\tau=1$)]{
			\includegraphics[width=8cm, height=6cm]{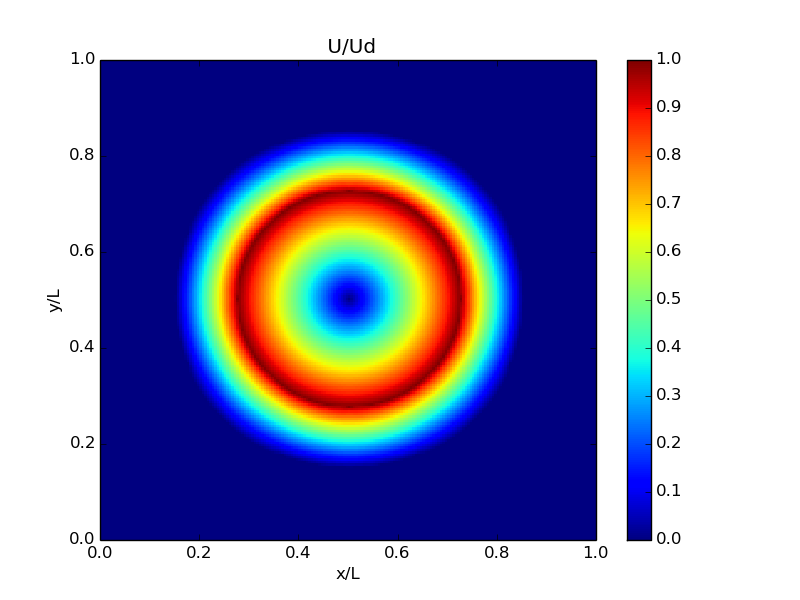}}     
		\hspace{-1.0cm}   
		\subfloat[VP SRT LBM ($\tau=10$)]{
			\includegraphics[width=8cm, height=6cm]{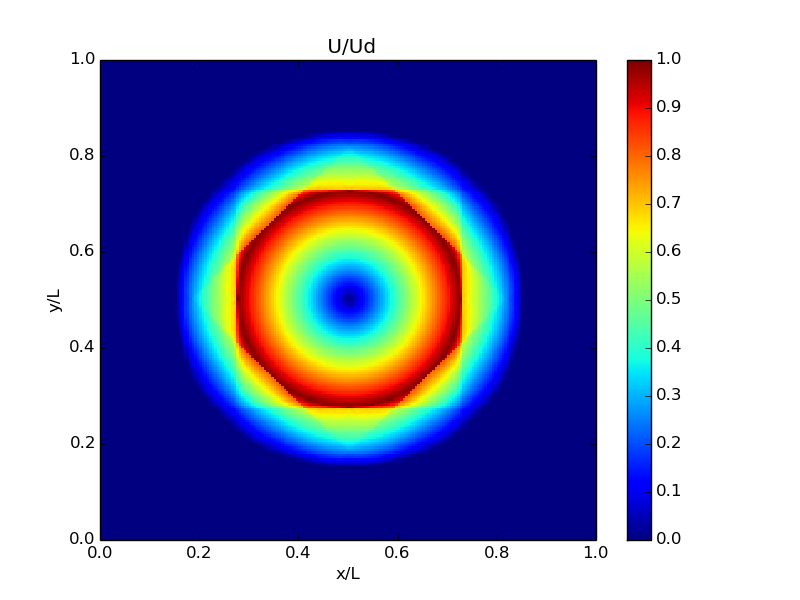}}        
		\vspace{-0.0cm}
		\subfloat[VP MRT LBM ($\tau=1$)]{
			\includegraphics[width=8cm, height=6cm]{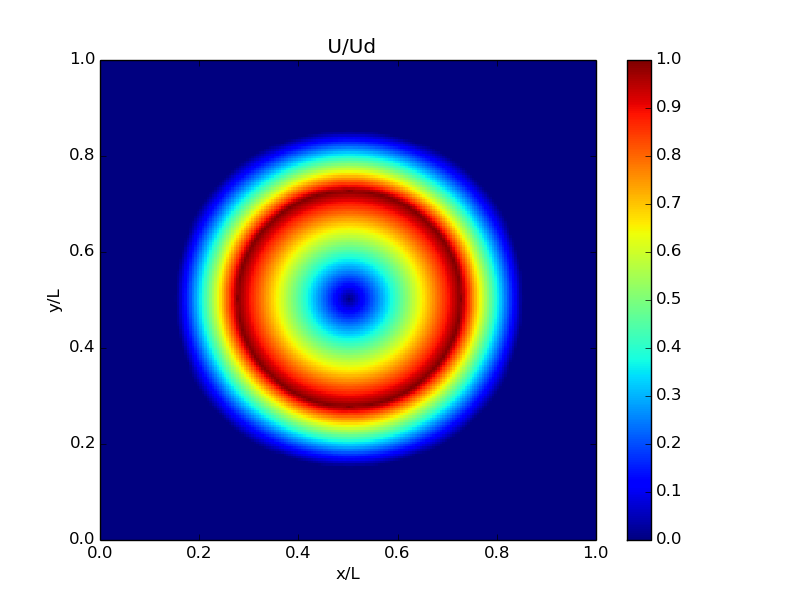}}      
		\hspace{-1.0cm}   
		\subfloat[VP MRT LBM ($\tau=10$)]{
			\includegraphics[width=8cm, height=6cm]{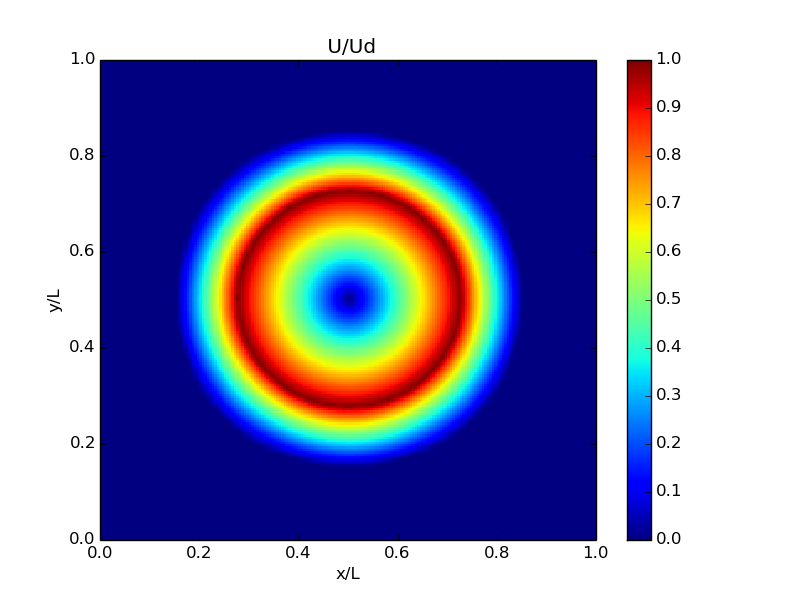}}            
		\caption{Isocontours of velocity magnitude obtained for the cylinder case. }
		\label{Cylindre2}
	\end{center}
\end{figure}
%%%%%%%%%%%%%%%%%%%%%%%%%%%%%%%%%%%%%%%%%%%%%%%%%%%%%%%%%%%%%%%%%%%%%%%%%%%%%%%%%%%%%%%%%%%%%%%%%%%%%%%%%%%%%%%%%%%%%%%%%%%%%%%%%%%%%%%%%%%%%%%%%%%%%%%%%%%%%%%
Figure 10 displays the isovalues of the velocity magnitude computed with the single relaxation time LBM and the multiple relaxation time LBM, 
for two relaxation times 
($\tau=1$ and $\tau=10$). In this figure, we can see that unrealistic results are obtained if a high relaxation time is chosen when the flow is computed with 
the single relaxation
time LBM ($\tau=10$, figure 10 (b)). This problem does not occur when the multiple relaxation time model is applied 
(see figure 10 (d)).\\
%%%%%%%%%%%%%%%%%%%%%%%%%%%%%%%%%%%%%%%%%%%%%%%%%%%%%%%%%%%%%%%%%%%%%%%%%%%%%%%%%%%%%%%%%%%%%%%%%%%%%%%%%%%%%%%%%%%%%%%%%%%%%%%%%%%%%%%%%%%%%%%%%%%%%%%%%%%%%%
\begin{figure}[h!]
	\centering
	\includegraphics[width=6.5cm,height=5cm, angle=0]{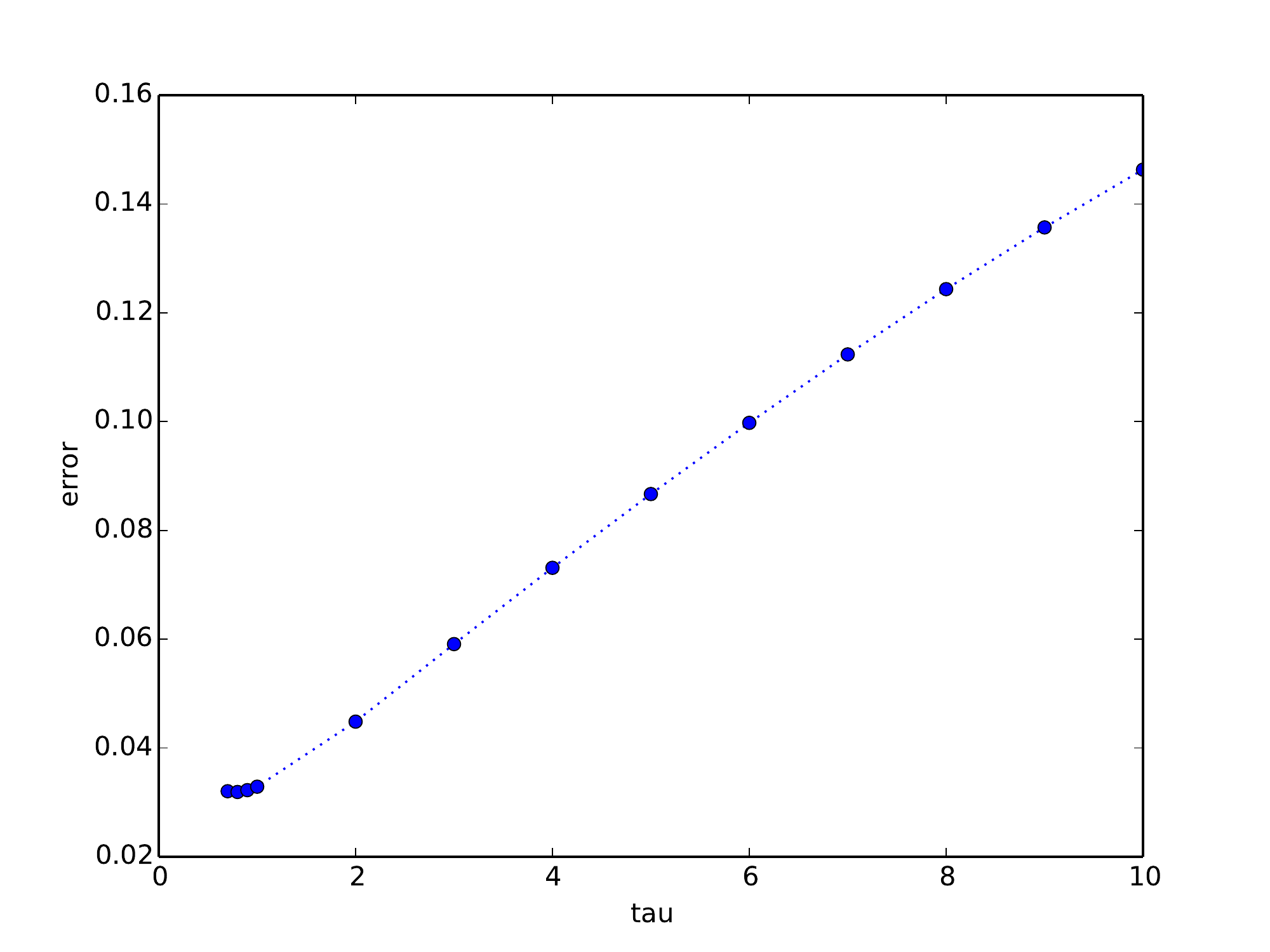}
	\caption{Influence of $\tau$ on the relative $L_2$ error of $v$, for the flow between two cylinders (VP SRT LBM).}
	\label{error_L2_cyl1}
\end{figure}
%%%%%%%%%%%%%%%%%%%%%%%%%%%%%%%%%%%%%%%%%%%%%%%%%%%%%%%%%%%%%%%%%%%%%%%%%%%%%%%%%%%%%%%%%%%%%%%%%%%%%%%%%%%%%%%%%%%%%%%%%%%%%%%%%%%%%%%%%%%%%%%%%%%%%%%%%%%%
\begin{figure}[h!]
  \begin{center}
   \leavevmode 
   \subfloat[influence of $\Lambda$ (VP MRT LBM)]{
     \includegraphics[width=6.5cm, height=5cm]{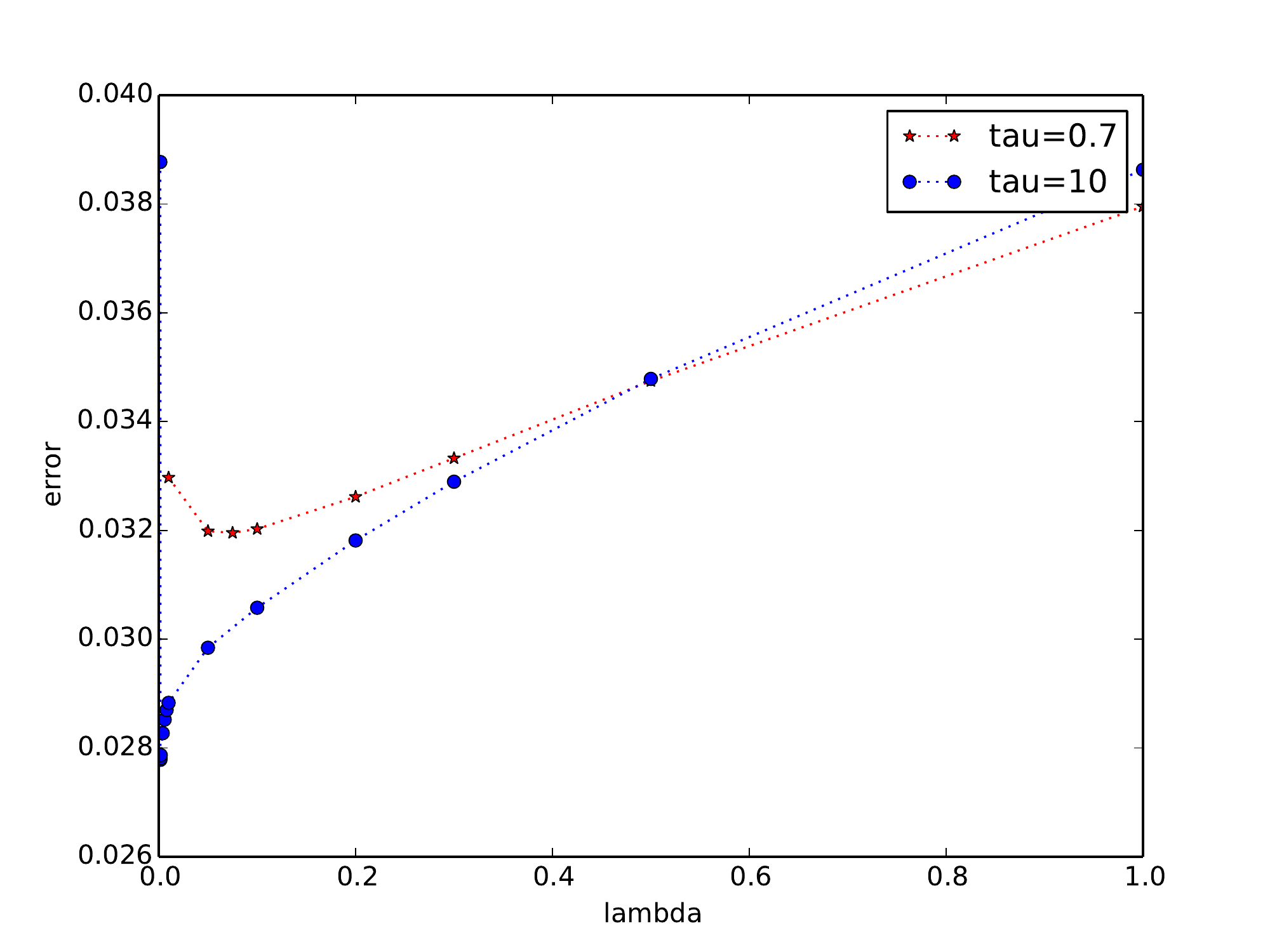}}     
    \hspace{-0.5cm}   
  \subfloat[influence of $\Lambda$ for $\tau=10$ : zoom on $\Lambda<0.06$]{
    \includegraphics[width=6.5cm, height=5cm]{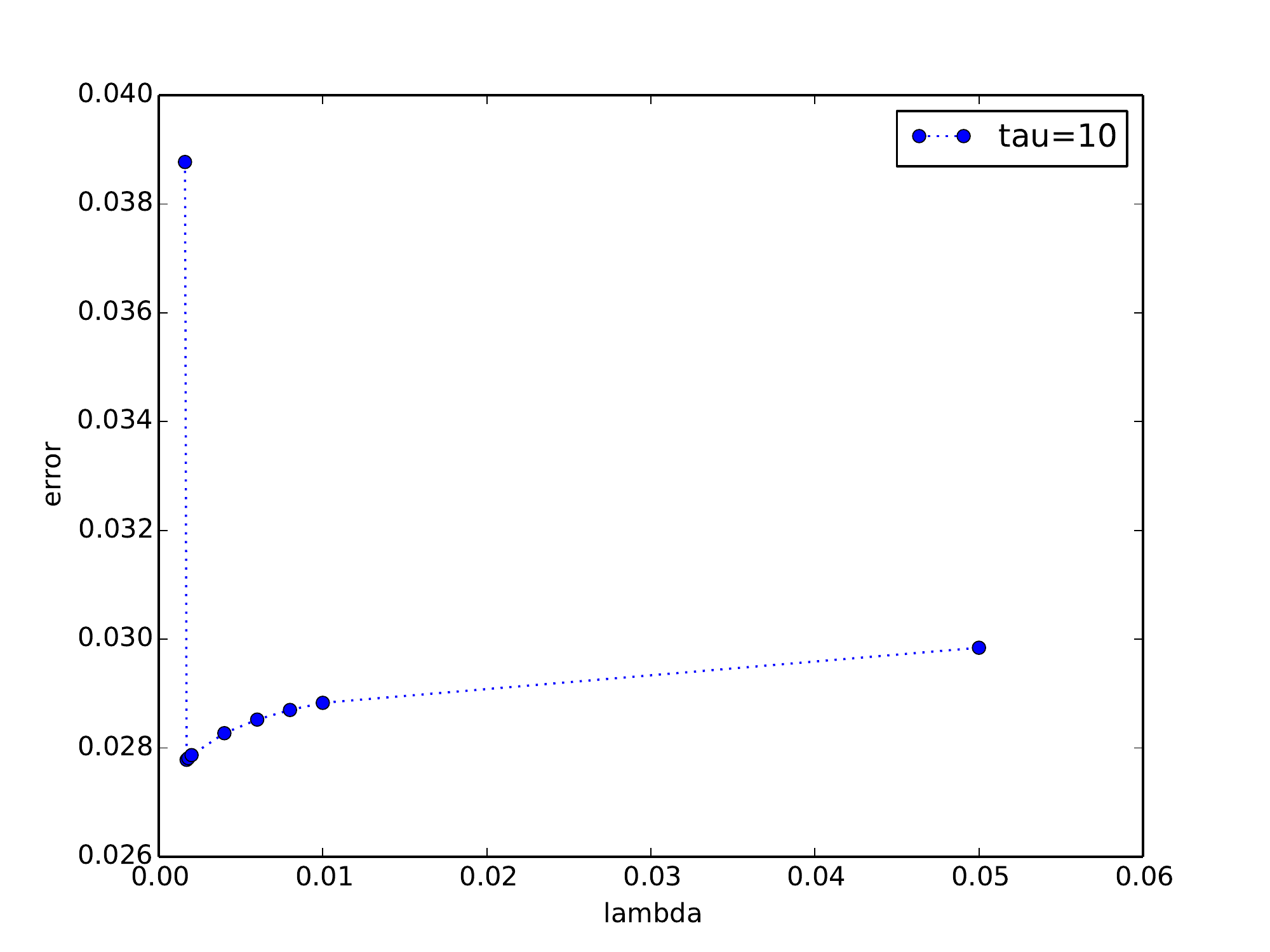}}        
     \vspace{-0.065cm}            
     \caption{(a) and (b) Influence of $\Lambda$ 
       on the relative $L_2$ error of $v$, for the flow between two cylinders (VP MRT LBM).}
    \label{error_L2_cyl2}
  \end{center}
\end{figure}
%%%%%%%%%%%%%%%%%%%%%%%%%%%%%%%%%%%%%%%%%%%%%%%%%%%%%%%%%%%%%%%%%%%%%%%%%%%%%%%%%%%%%%%%%%%%%%%%%%%%%%%%%%%%%%%%%%%%%%%%%%%%%%%%%%%%%%%%%%%%%%%%%%%%%%%%%%%%%%%
\begin{figure}[h!]
	\begin{center}
		\leavevmode 
		\subfloat[difference between the numerical (VP SRT LBM) and the exact vertical velocities]{
			\includegraphics[width=6.5cm, height=5cm]{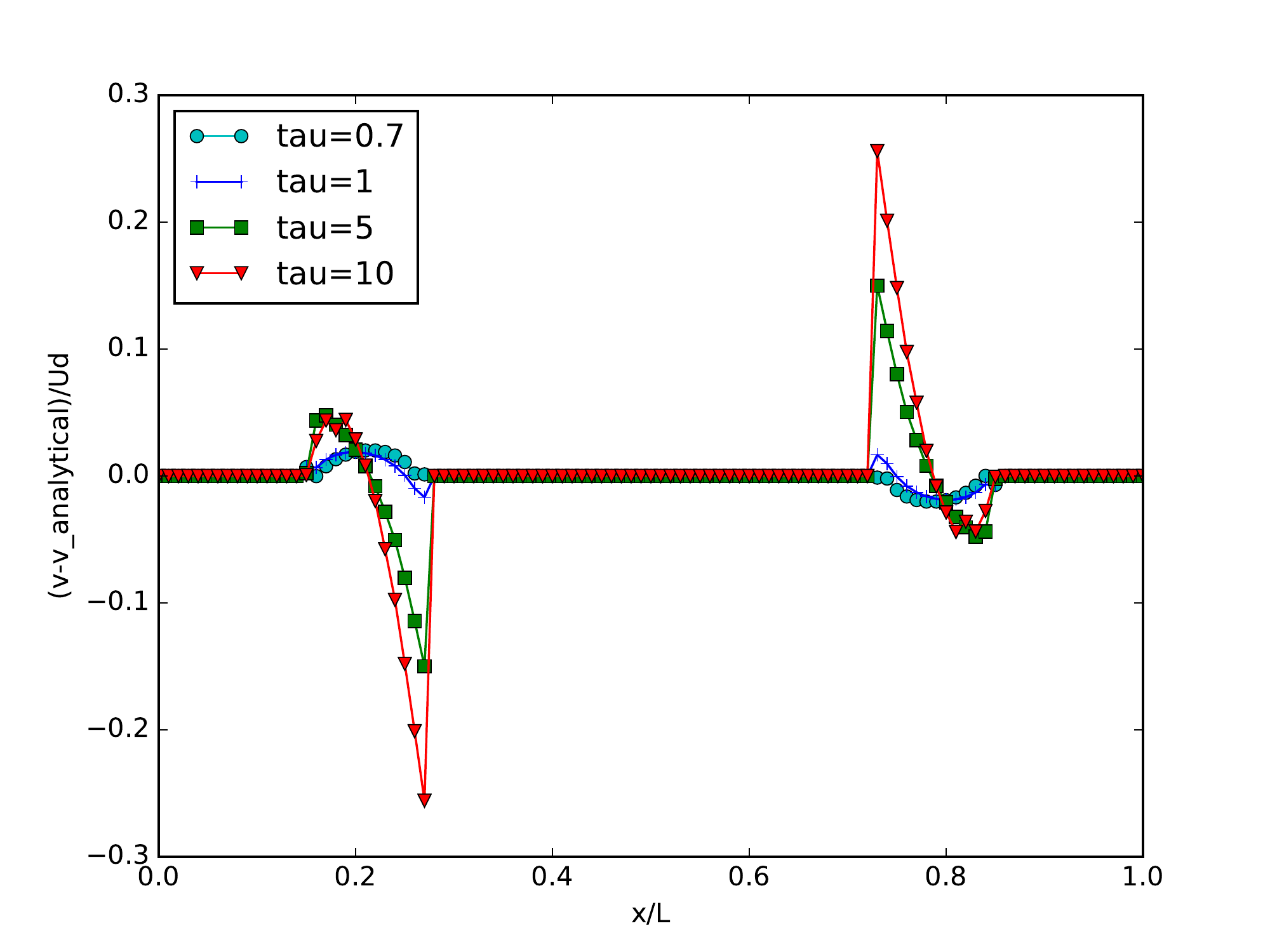}}     
		\hspace{0.5cm}   
		\subfloat[vertical velocity obtained with the VP SRT LBM]{
			\includegraphics[width=6.5cm, height=5cm]{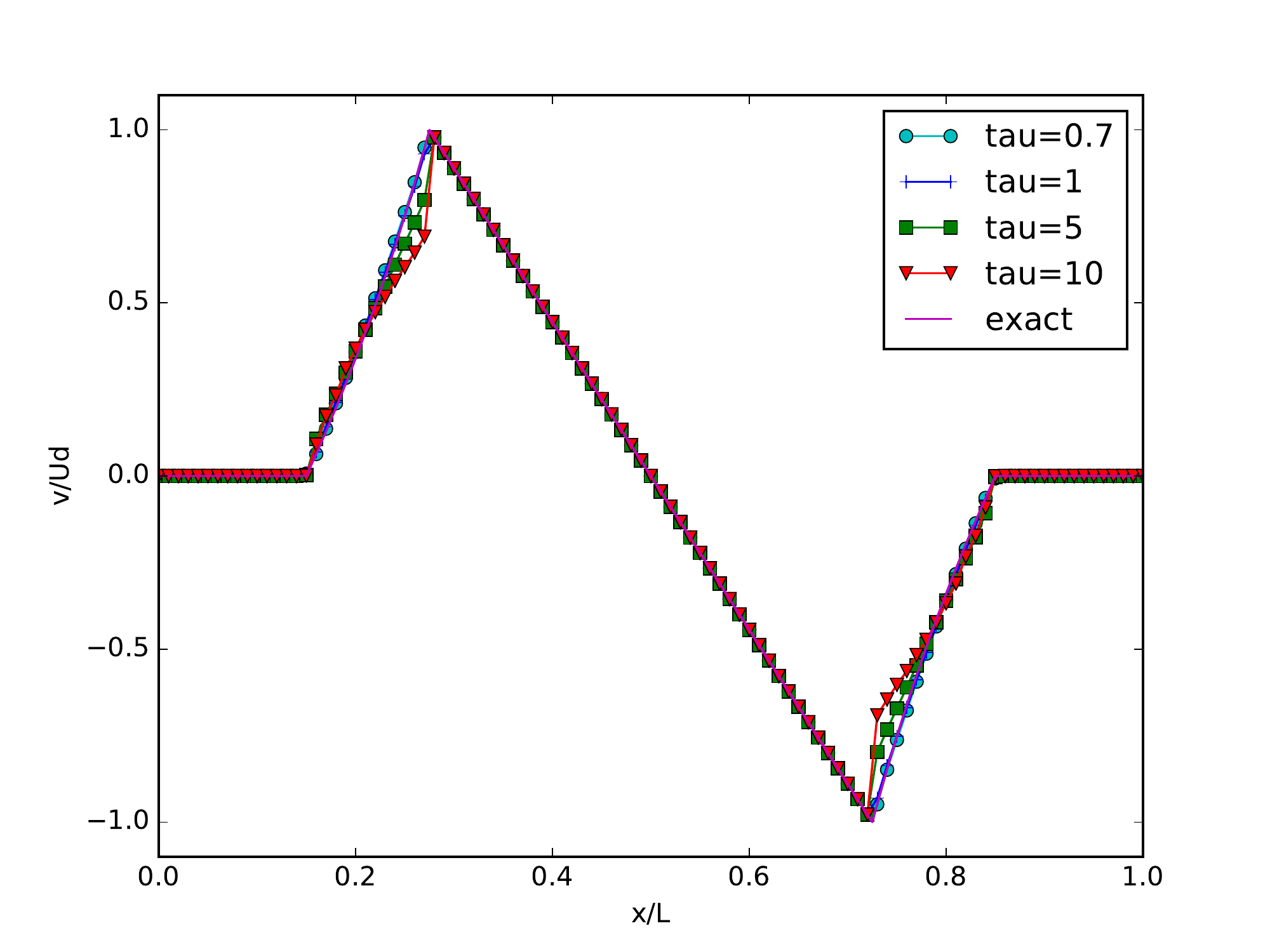}}        
		\vspace{-0.065cm}
		\subfloat[difference between the numerical (VP MRT LBM) and the exact vertical velocities]{
			\includegraphics[width=6.5cm, height=5cm]{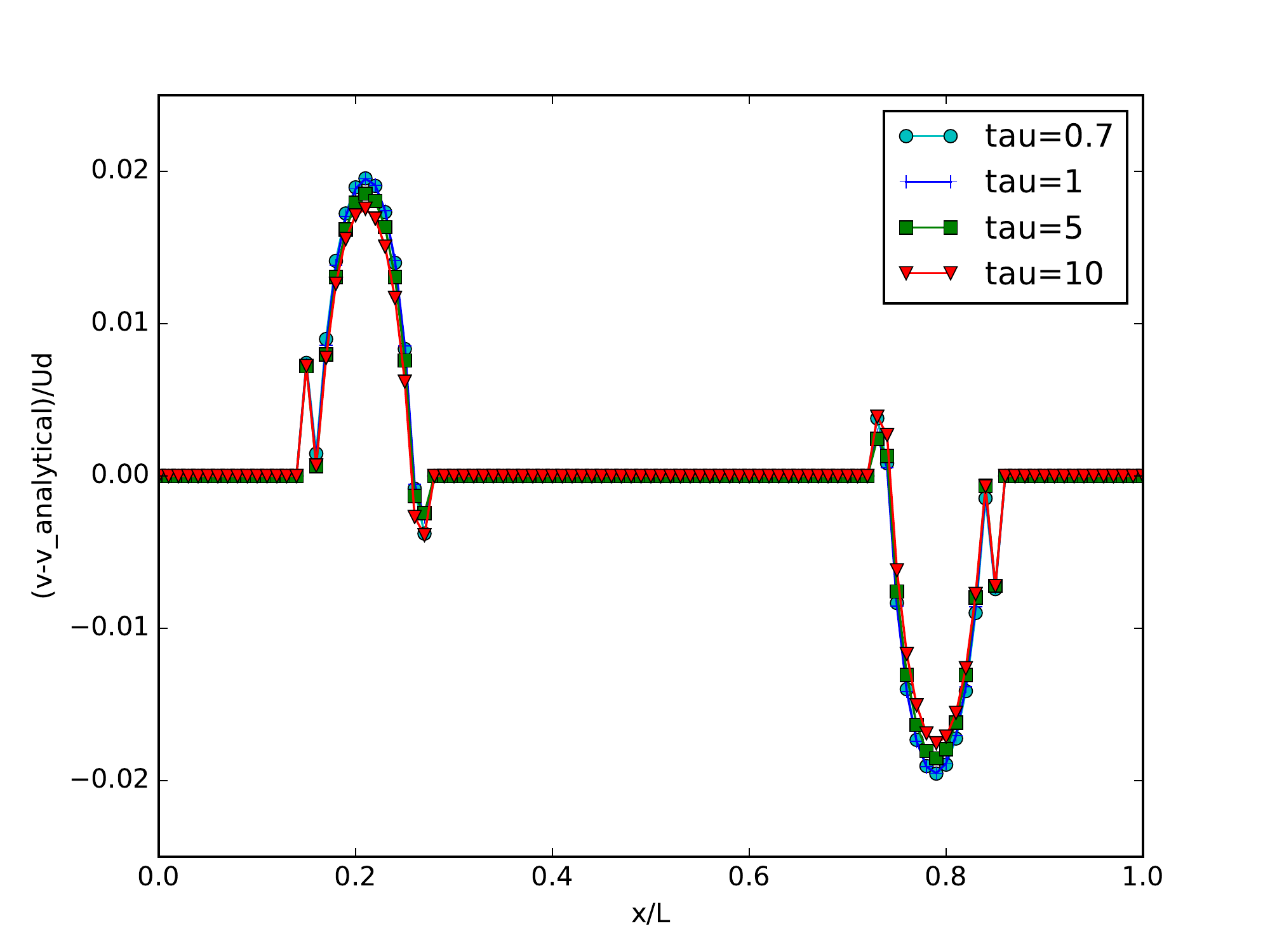}}      
		\hspace{0.5cm}   
		\subfloat[vertical velocity obtained with the VP MRT LBM]{
			\includegraphics[width=6.5cm, height=5cm]{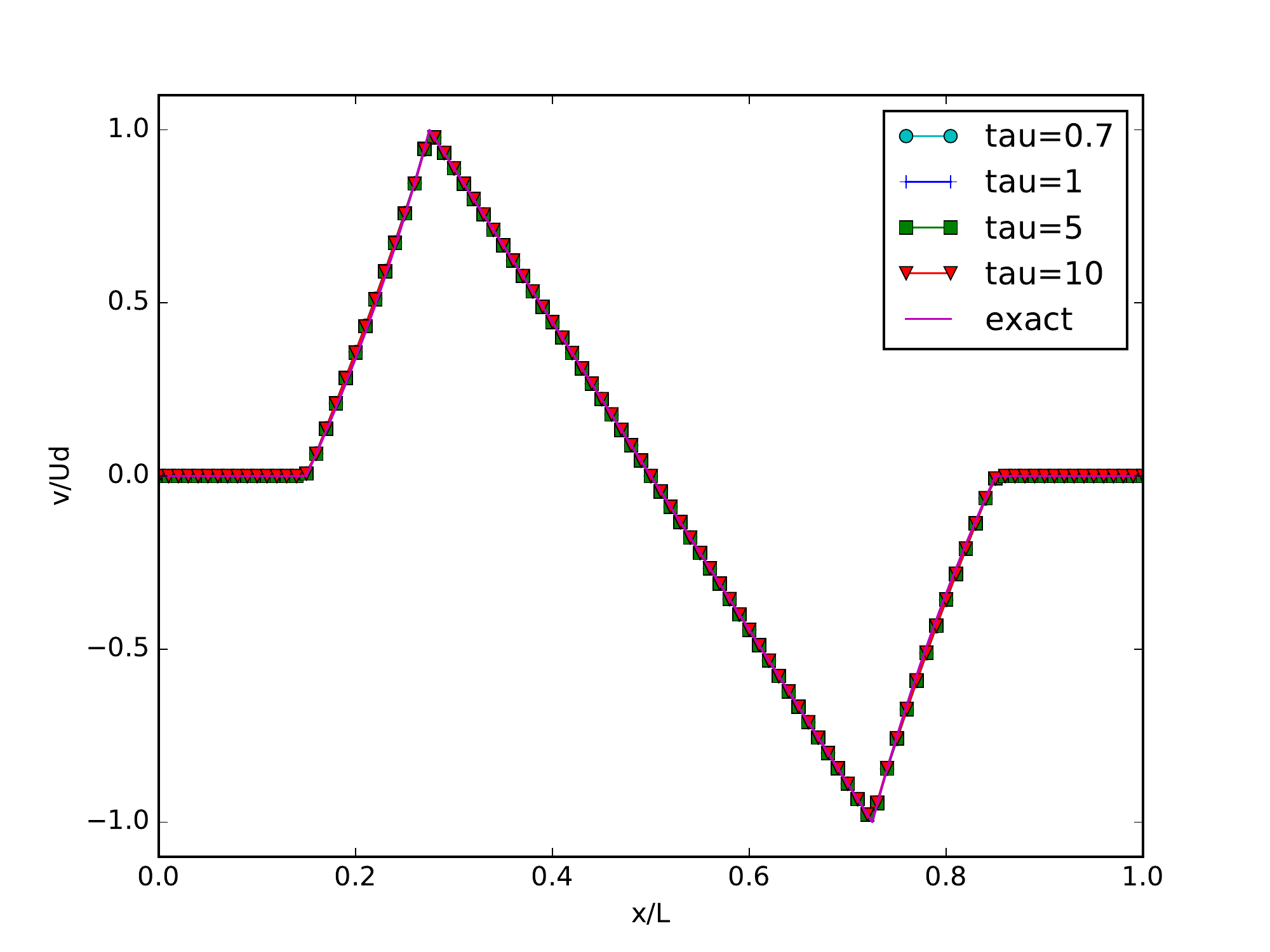}}            
		\caption{Difference between the numerical and the exact velocities, and velocity profiles, obtained at $y = L/2$, for the cylinder case.}
		\label{Cylindre3}
	\end{center}
\end{figure}
%%%%%%%%%%%%%%%%%%%%%%%%%%%%%%%%%%%%%%%%%%%%%%%%%%%%%%%%%%%%%%%%%%%%%%%%%%%%%%%%%%%%%%%%%%%%%%%%%%%%%%%%%%%%%%%%%%%%%%%%%%%%%%%%%%%%%%%%%%%%%%%%%%%%%%%%%%%%%%%
In figure 11, the relative $L_2$ error of $v$ in the fluid domain as a function of the relaxation time, computed with the single relaxation 
time LBM, is plotted. In contrast 
with the previous cases which involved straight boundaries, this case deals with a curved boundary which is approximated in a staircase manner; 
this induces an error for all $\tau$
values. A similar remark can be done when examining the evolution of the relative $L_2$ error of $v$ according to $\Lambda$, for the computations performed 
with the multiple relaxation time LBM (see 
figure 12 (a)) and a relaxation time $\tau=0.7$. For a higher relaxation time ($\tau=10$), the optimal value 
of $\Lambda$ that minimizes the error is closer to zero, and the error is smaller than for $\tau=0.7$
(see figures 12 (a) and (b)).\\
%%%%%%%%%%%%%%%%%%%%%%%%%%%%%%%%%%%%%%%%%%%%%%%%%%%%%%%%%%%%%%%%%%%%%%%%%%%%%%%%%%%%%%%%%%%%%%%%%%%%%%%%%%%%%%%%%%%%%%%%%%%%%%%%%%%%%%%%%%%%%%%%%%
\begin{figure}[h!]
	\centering
			\includegraphics[width=6.5cm, height=5cm]{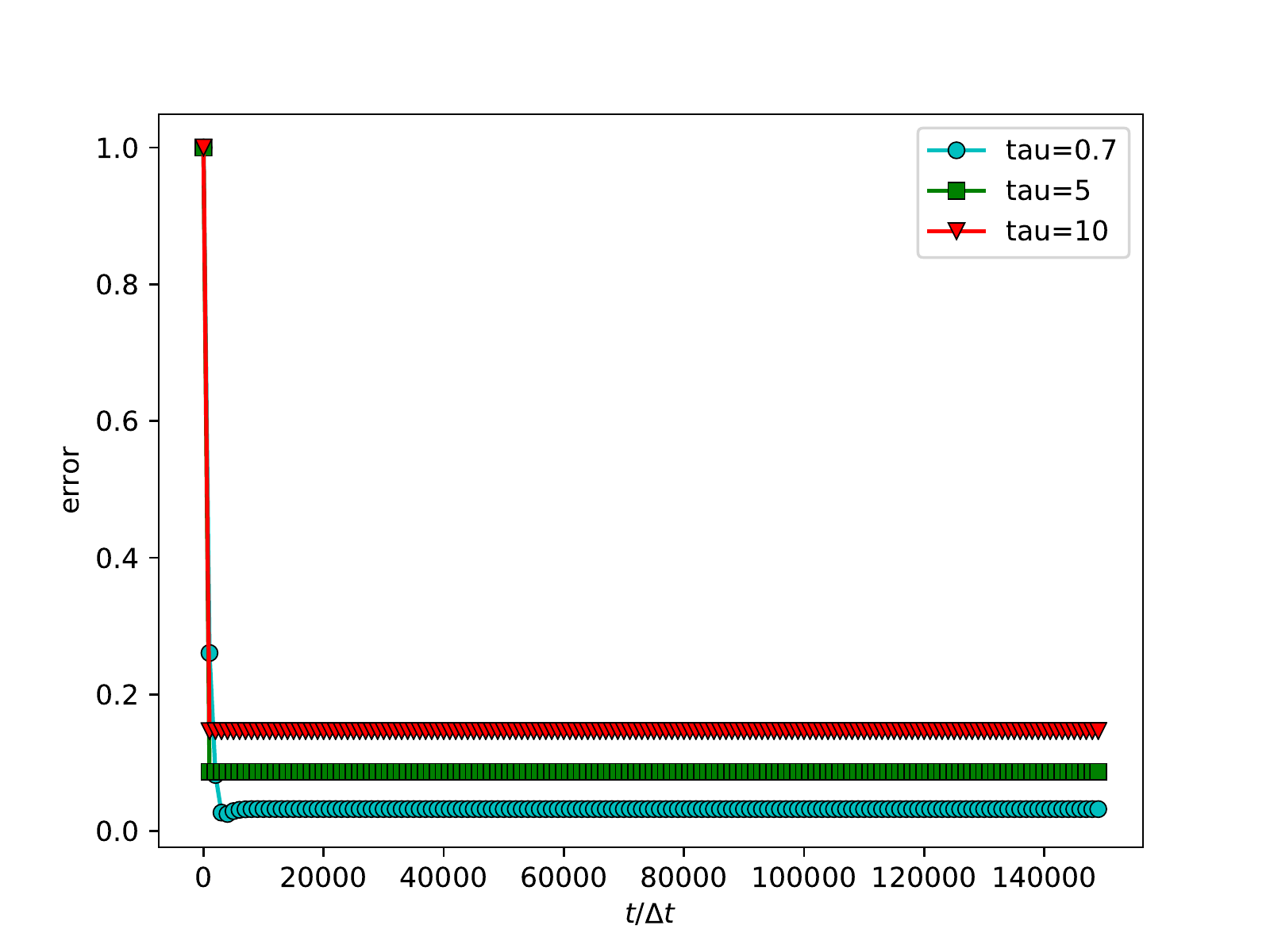}     
		\caption{Convergence history of the VP SRT LBM, for the cylinder case.}
		\label{histoire_globale_cylindre}
\end{figure}
%%%%%%%%%%%%%%%%%%%%%%%%%%%%%%%%%%%%%%%%%%%%%%%%%%%%%%%%%%%%%%%%%%%%%%%%%%%%%%%%%%%%%%%%%%%%%%%%%%%%%%%%%%%%%%%%%%%%%%%%%%%%%%%%%%%%%%%%%%%%%%%%%%
\begin{figure}[h!]
	\begin{center}
		\leavevmode 
		\subfloat[$\tau=0.7$]{
			\includegraphics[width=6.5cm, height=5cm]{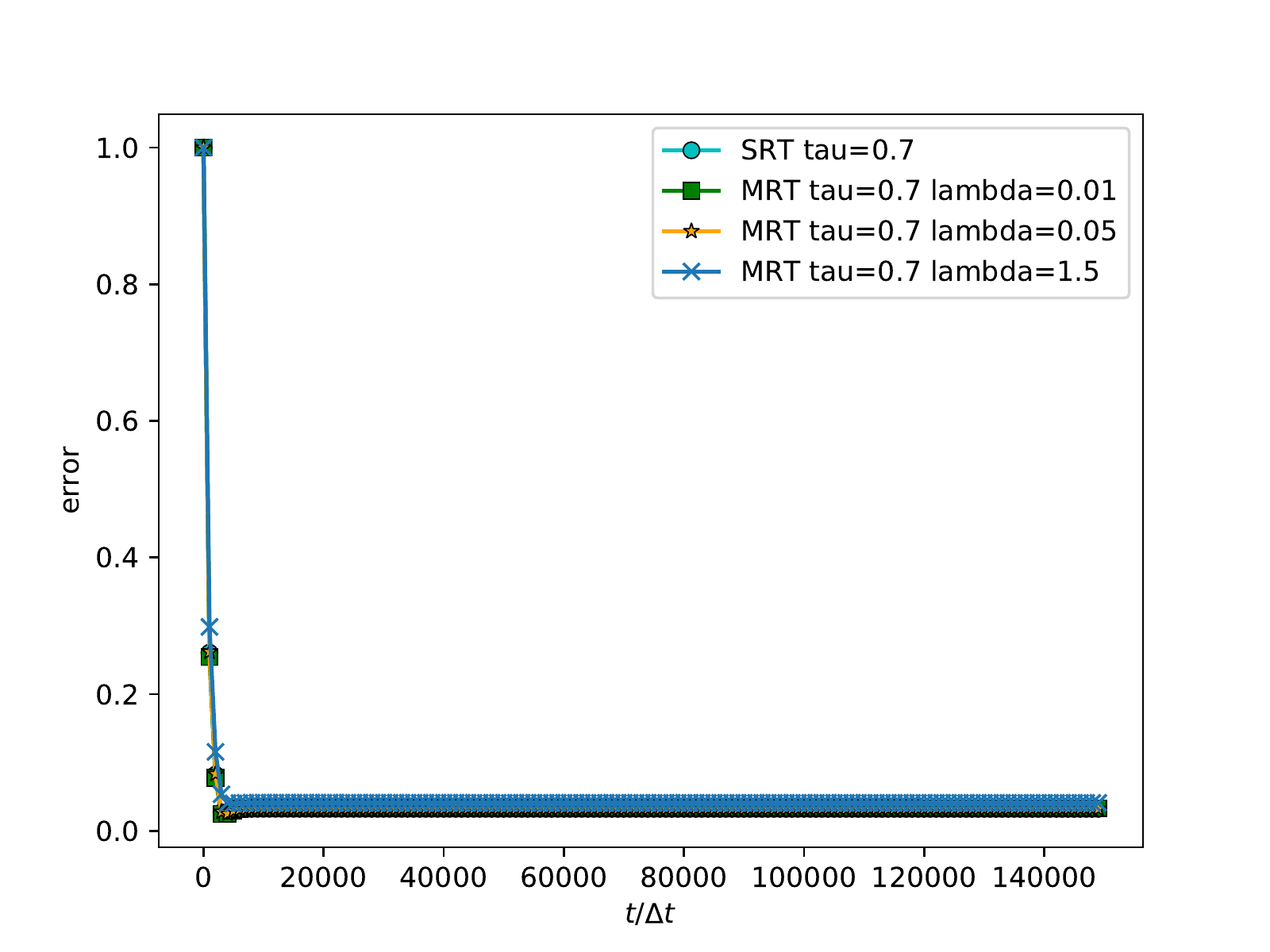}}     
		\hspace{0.5cm}   
		\subfloat[$\tau=10$]{
			\includegraphics[width=6.5cm, height=5cm]{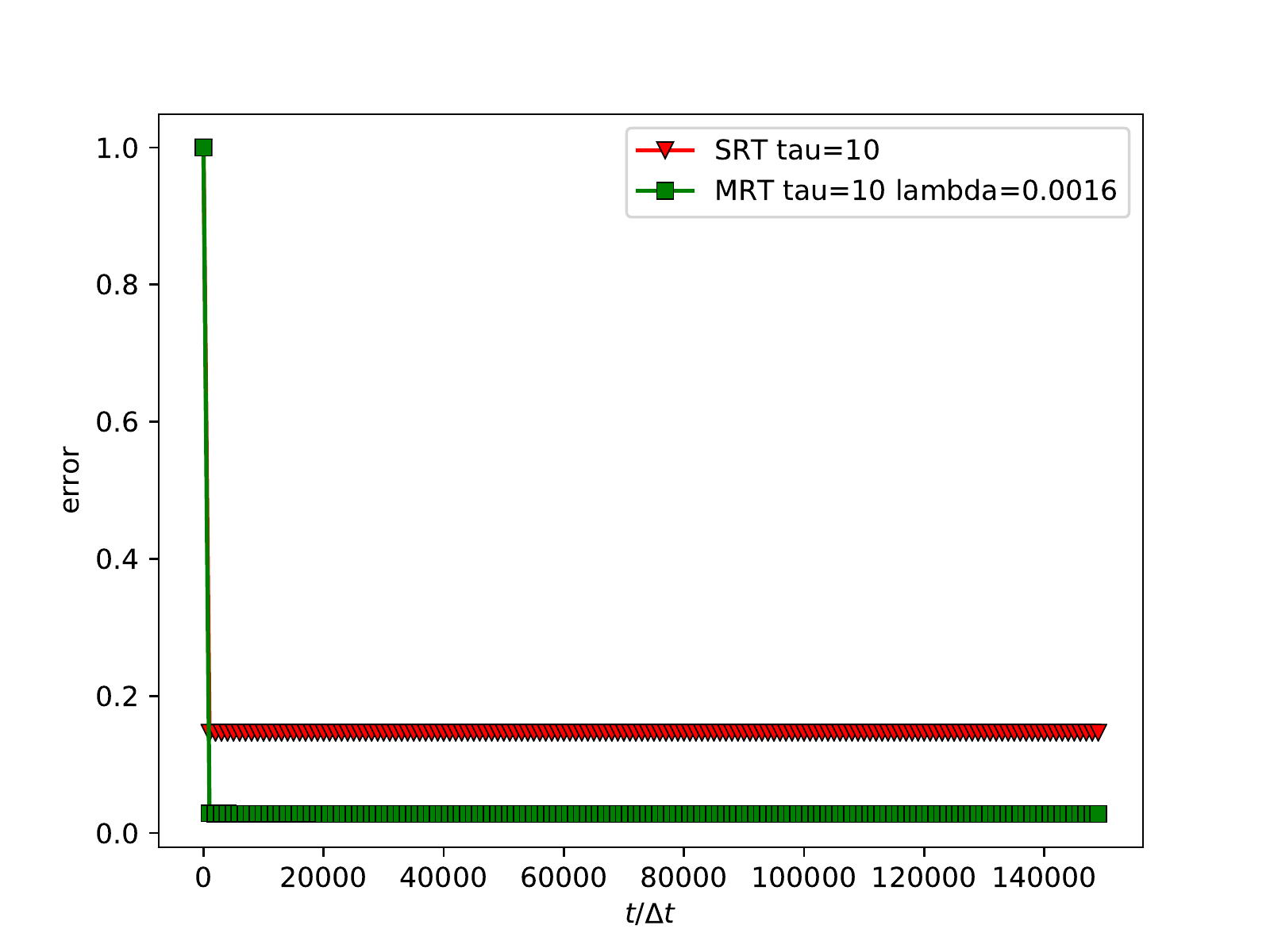}}        
		\caption{Convergence history of the VP MRT LBM, for the cylinder case.}
		\label{histoire_cylindre_MRT}
	\end{center}
\end{figure}
%%%%%%%%%%%%%%%%%%%%%%%%%%%%%%%%%%%%%%%%%%%%%%%%%%%%%%%%%%%%%%%%%%%%%%%%%%%%%%%%%%%%%%%%%%%%%%%%%%%%%%%%%%%%%%%%%%%%%%%%%%%%%%%%%%%%%%%%%%%%%%%%%%
Figures 13 (a) and (c) show the profiles of the difference (made non dimensional with $u^d$) between the numerical and the exact solutions, 
obtained with the single relaxation time LBM, 
and with the multiple relaxation time LBM, at $y=L/2$. The corresponding velocity profiles are plotted in figures 13 (b) and (d). 
The optimal values of $\Lambda$ that minimize the relative $L_2$ error of $v$ were used for the computations performed with the MRT LBM.
In this figure, we
notice that, for high values of the relaxation time, the volume penalization MRT model produces errors that are smaller than the ones obtained 
with the volume penalization SRT model. High differences can also
be seen at the fluid-solid interface for the VP SRT LBM and high values of the relaxation time. For this case with curved boundaries, we can also
remark that when applying the MRT LBM with $\tau=10$ and $\Lambda=0.0016$, converged results are obtained after a smaller number of iterations than
with other relaxation times (for the MRT LBM, and the SRT LBM, see figures 14 and 15). 
However, the discrepancy between $\tau=0.7$ and $\tau=10$ for the MRT LBM is not high.\\
In figure 16, 
we also compared the errors obtained with
 the volume penalization LBM (SRT and MRT) with those computed with the immersed boundary method (and the 
 same grid) by Lu et al. \cite{lu_immersed_2012}.  
 In this figure, we can notice that the error in the vicinity of the solid boundary is higher with the volume penalization method that approximates the
 curved boundary with a staircase manner, than with the immersed boundary method that takes the curved shape of the boundary into consideration. The 
 error in the rest of the fluid domain is lower when the flow is predicted with the VP LBM.\\
%%%%%%%%%%%%%%%%%%%%%%%%%%%%%%%%%%%%%%%%%%%%%%%%%%%%%%%%%%%%%%%%%%%%%%%%%%%%%%%%%%%%%%%%%%%%%%%%%%%%%%%%%%%%%%%%%%%%%%%%%%%%%%%%%%%%%%%%%%%%%%%%%%%
\begin{figure}[h!]
	\begin{center}
		\leavevmode 
		\subfloat[SRT LBM]{
			\includegraphics[width=6.5cm, height=5cm]{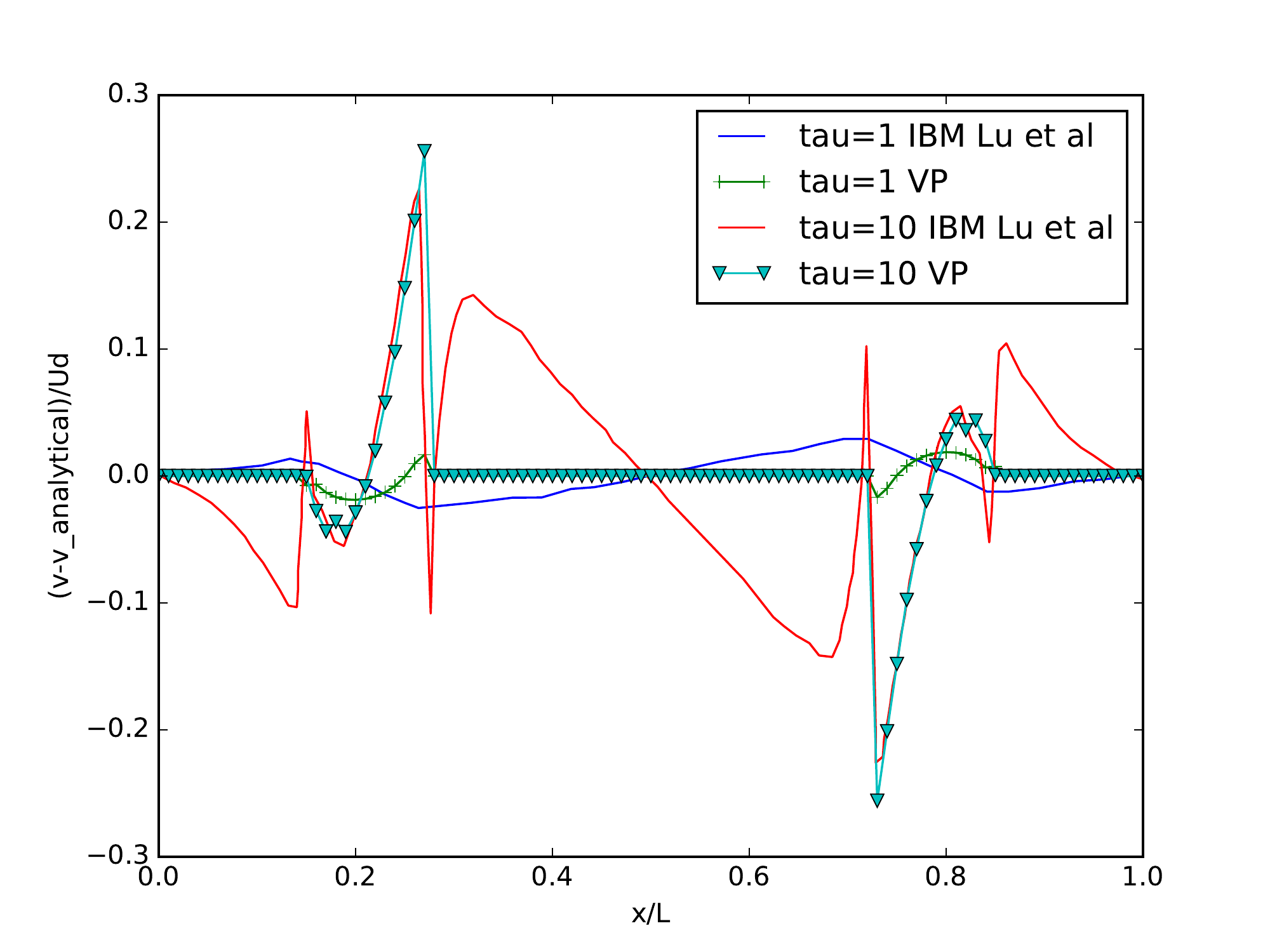}}     
		\hspace{0.5cm}   
		\subfloat[MRT LBM]{
			\includegraphics[width=6.5cm, height=5cm]{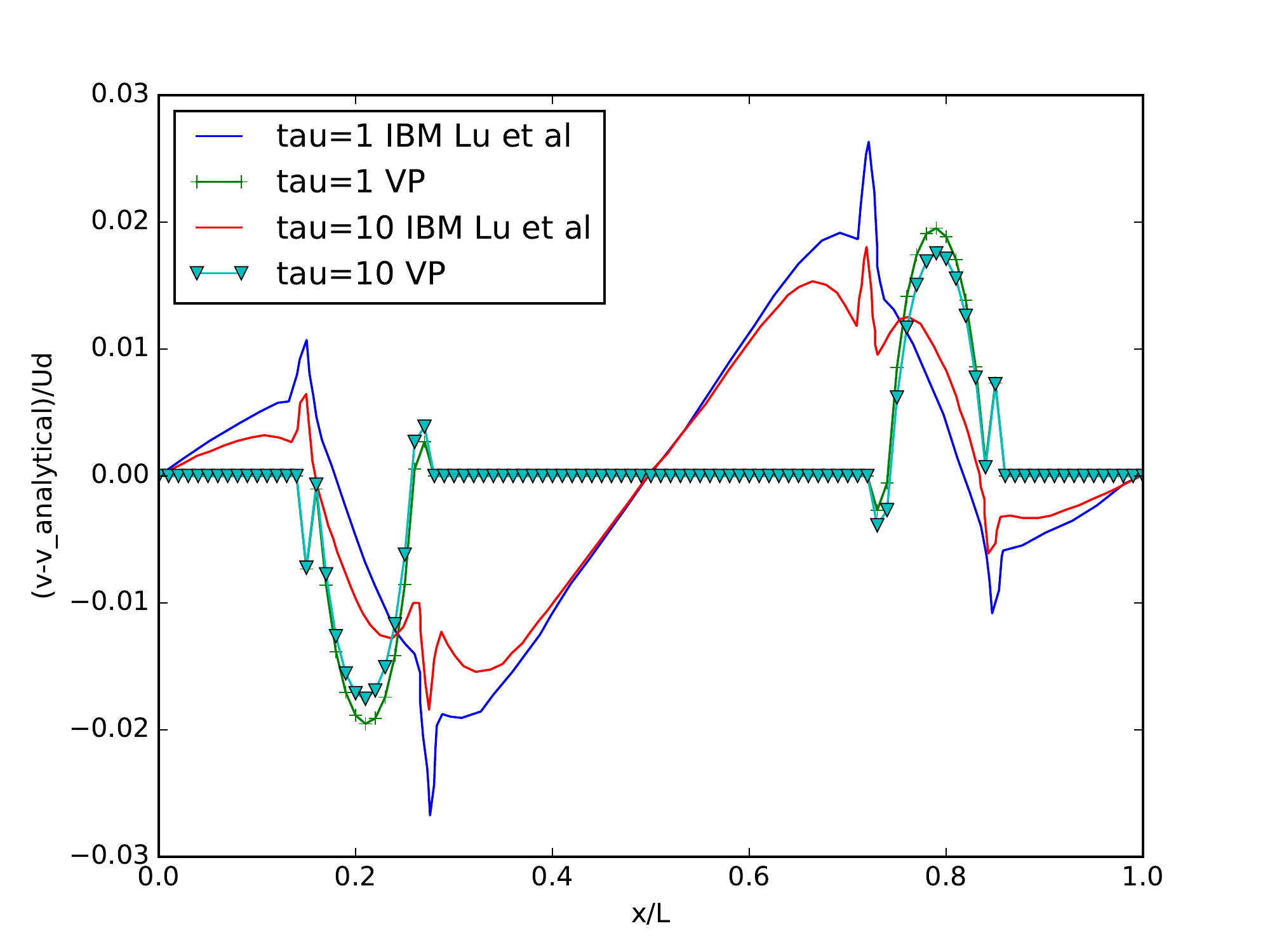}}        
		\caption{Comparison of the errors obtained with the volume penalization LBM, with those computed with an immersed boundary LBM by Lu et 
		al. \cite{lu_immersed_2012}, for the cylinder case.}
		\label{comparaison_guo_cyl}
	\end{center}
\end{figure}
%%%%%%%%%%%%%%%%%%%%%%%%%%%%%%%%%%%%%%%%%%%%%%%%%%%%%%%%%%%%%%%%%%%%%%%%%%%%%%%%%%%%%%%%%%%%%%%%%%%%%%%%%%%%%%%%%%%%%%%%%%%%%%%%%%%%%%%%%%%%%%%%%%
For this case, 
we also studied the influence of the grid on the accuracy of the results. To do that, we carried out simulations with the multiple relaxation time LBM, 
and for a relaxation time $\tau=0.7$. The number of cells according to the x and y directions was varied between $200 \;l.u.$ and $800 \;l.u.$ 
For the dense grid ($800\; l.u.$), two cases were considered: one case where the number of grids Ngrid in the flow domain, according to x and y, 
was: $R_2-R_1=100 \;l.u.$, and another case where this number was higher: $R_2-R_1=125 \;l.u.$ (see table 1). In 
figure 17, the evolution of the relative $L_2$ error of $v$ according to the parameter $\Lambda$ is presented. 
We can see that for a higher number of cells in the flow domain, the error decreases and the results are less dependent on the 
parameter $\Lambda$. To accurately describe the curved surface with the volume penalization method, a large number of grids must thus be used.
%%%%%%%%%%%%%%%%%%%%%%%%%%%%%%%%%%%%%%%%%%%%%%%%%%%%%%%%%%%%%%%%%%%%%%%%%%%%%%%%%%%%%%%%%%%%%%%%%%%%%%%%%%%%%%%%%%%%%%%%%%%%%%%%%%%%%%%%%%%%%%%%%%%%
\begin{table}[h!]
	\begin{center}
		\begin{tabular}{|cc|}
			\hline
			L ($l.u.$) & Ngrid ($l.u.$)\\
			
			200 & 25 \\
			
			400 & 50 \\ 
			
			600 & 75 \\ 
			
			800 & 100 \\ 
			
			800 & 125 \\ 
			\hline	
		\end{tabular}
	\end{center}
	\caption{Grid resolutions tested for the cylinder case}	
	\label{maille_cyl}
\end{table}

\begin{figure}[h!]
	\centering
	\includegraphics[width=6.5cm,height=5cm, angle=0]{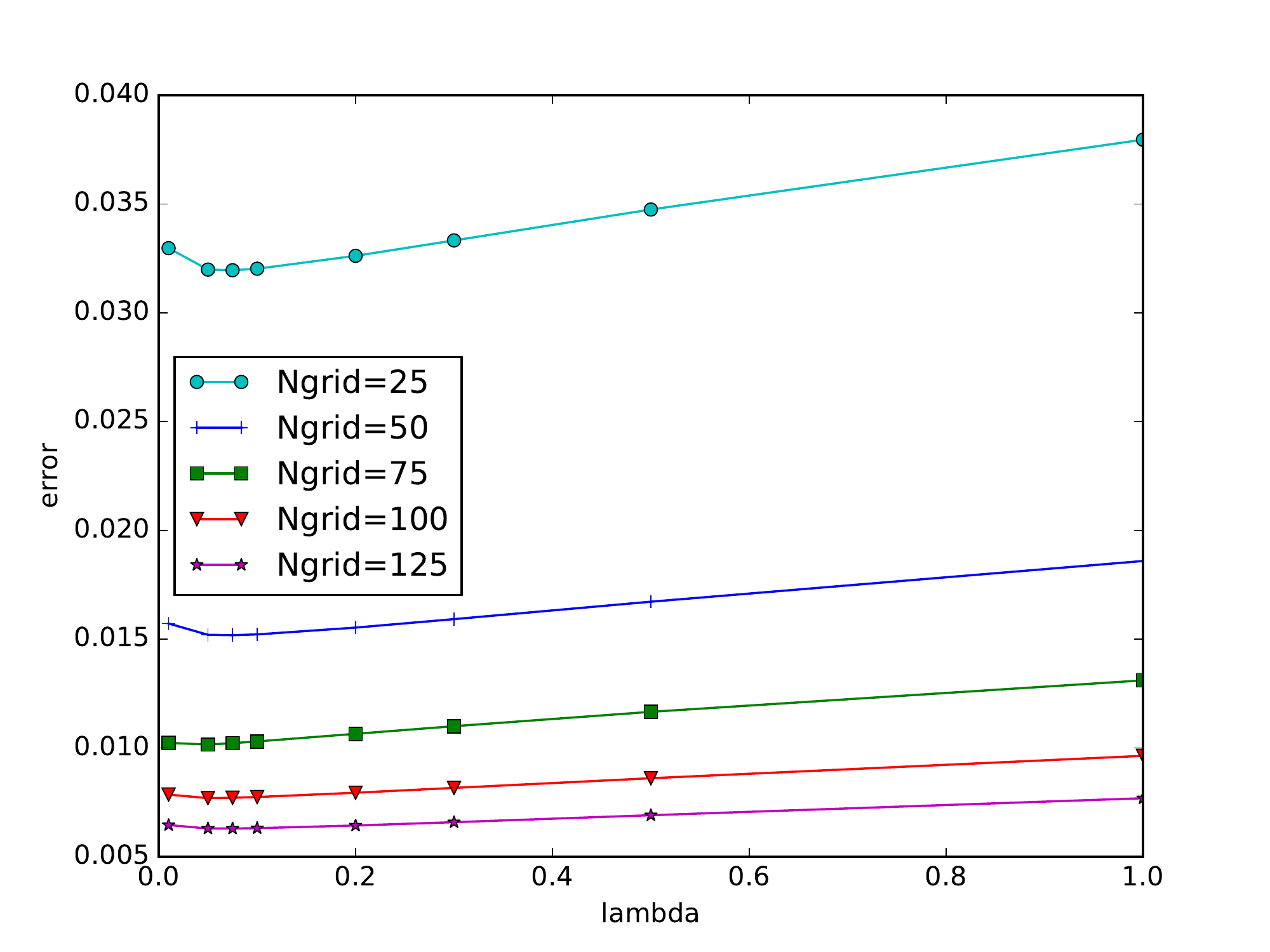}
	\caption{Influence of the grid resolution on the relative $L_2$ error of $v$, for the flow between two cylinders (VP MRT LBM, $\tau=0.7$).}
	\label{error_L2_cyl_grid1}
\end{figure}
%%%%%%%%%%%%%%%%%%%%%%%%%%%%%%%%%%%%%%%%%%%%%%%%%%%%%%%%%%%%%%%%%%%%%%%%%%%%%%%%%%%%%%%%%%%%%%%%%%%%%%%%%%%%%%%%%%%%%%%%%%%%%%%%%%%%%%%%%%%%%%%%%%%%%%%
\subsection{Flow past a circular cylinder at a Reynolds number of 500}
The cases presented in the previous paragraphs deal with low Reynolds number flows. In our previous paper, we applied the VP LBM method to the computation of the 
flow past a cylinder at a Reynolds number of 100. In this paper, to evaluate 
the ability of the VP LBM to handle more complex flows, we focused on the flow around a circular cylinder at a Reynolds number of 500 (the Reynolds number is based on the diameter $D$ of the cylinder 
and the free stream velocity $U_{fs}$). The computational domain is presented in figure 18. For that case, the boundary conditions at the upper and lower horizontal
planes were symmetry boundary conditions, a constant velocity profile was applied at the inlet, and at the outlet, a convective condition, as proposed
by Yang \cite{Yang_2013}, was employed. The dimensions of the domain are $30D$ in the horizontal direction, $10D$ in the vertical direction, and the distance
between the cylinder and the outlet is equal to $25D$.\\
As highlighted in the previous paragraph, a large number of grids must be used to describe properly a curved surface, when using the volume penalization method.
Hence, the domain was discretized with 6048 lattices according to the horizontal direction, 2016 lattices according
to the vertical direction, and 201 lattices along the diameter of the cylinder. A velocity of 0.05 was imposed at the inlet. For that case which involved 
a large number of grids, the VP-LBM code was parallelized with CUDA. The computations were carried out with Graphics Processing Units (GPUs), and the single relaxation time lattice Boltzmann method
was used. A non dimensional relaxation time $\tau=0.56$ was chosen.\\
In figure 19 which displays the instantaneous isovalues of velocity magnitude around the cylinder, we can see the vortex shedding phenomenon 
that occurs behind the cylinder. With the momentum exchange method \cite{Ladd_1994}, the drag force $F_D$ and the lift force $F_L$
exerted on the cylinder were calculated, and the drag and lift coefficients $C_D$ and $C_L$ were obtained from the following relationships:
\begin{equation}
 C_D=\frac{F_D}{\frac{1}{2}\rho U_{fs}^2 D} \text{       and      } C_L=\frac{F_L}{\frac{1}{2}\rho U_{fs}^2 D}
\end{equation}
The Strouhal number was also calculated as follows:
\begin{equation}
 St=\frac{f D}{U_{fs}}
\end{equation}
where $f$ is the frequency of the vortex shedding. In figure 20, where the evolutions of the drag and the lift coefficients according to the non dimensional time 
are presented, a periodic behavior of these coefficients can be noticed.
The average drag coefficient $C_{Dave}$, the root mean square of the lift coefficient $C_{Lrms}$, and the 
Strouhal number were compared with experimental results found in the literature \cite{Wen_2001, Wen_2004}, and with numerical results obtained with the 
finite volume method (FVM), and more particularly with the OpenFOAM \cite{OpenFOAM} code, on a structured mesh. In table 2, we can see that the results
calculated with the VP-LBM are in a satisfactory agreement with experimental results, and with numerical results computed with the finite volume method.
%%%%%%%%%%%%%%%%%%%%%%%%%%%%%%%%%%%%%%%%%%%%%%%%%%%%%%%%%%%%%%%%%%%%%%%%%%%%%%%%%%%%%%%%%%%%%%%%%%%%%%%%%%%%%%%%%%%%%%%%%%%%%%%%%%%%%%%%%%%%%%%%%%%%%%
\begin{figure}[h!]
	\centering
	\includegraphics[width=10.5cm,height=5cm, angle=0]{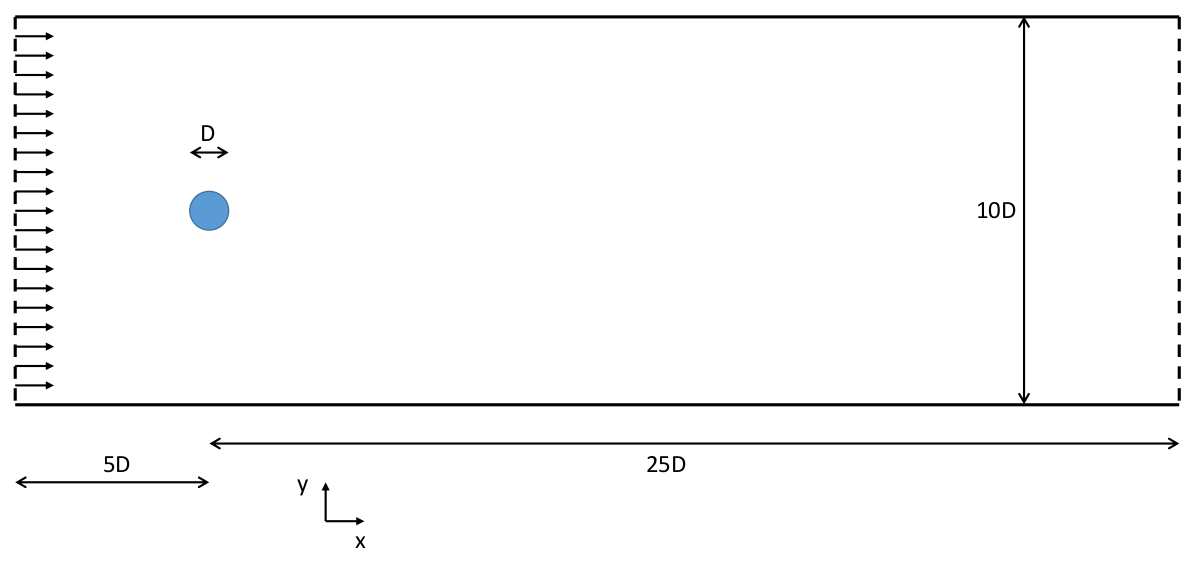}
	\caption{Computational domain for the flow around a cylinder.}
	\label{cylindre_dans_canal}
\end{figure}

\begin{figure}[h!]
	\centering
	\includegraphics[width=10.5cm,height=5cm, angle=0]{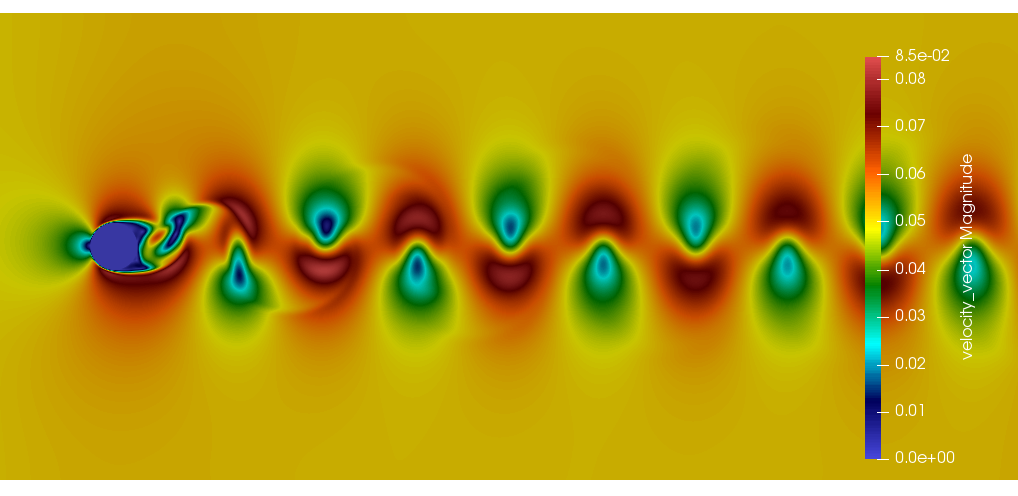}
	\caption{Instantaneous isovalues of velocity magnitude behind the cylinder ($Re=500$).}
	\label{allees_Von_Karman}
\end{figure}

\begin{figure}[h!]
	\begin{center}
		\leavevmode 
		\subfloat[Drag coefficient]{
			\includegraphics[width=6.5cm, height=5cm]{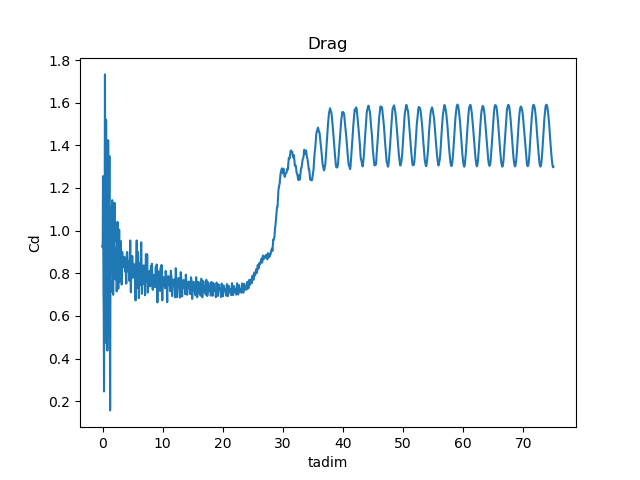}}     
		\hspace{0.5cm}   
		\subfloat[Lift coefficient]{
			\includegraphics[width=6.5cm, height=5cm]{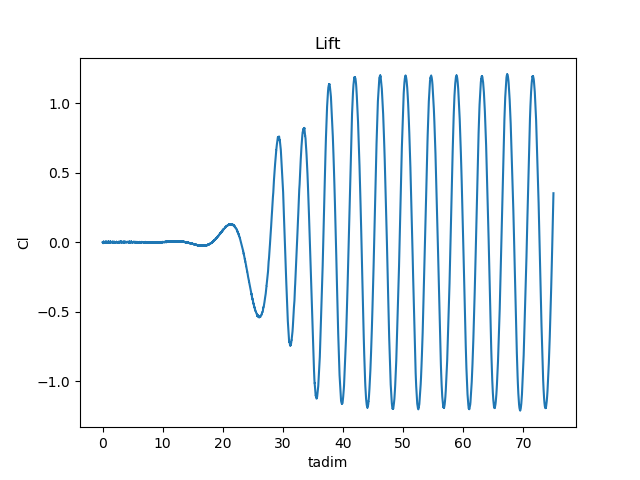}}        
		\caption{Drag and lift coefficients obtained with the VP LBM,
		for the flow past a circular cylinder, at a Reynolds number of 500.}
		\label{drag_lift_cylinder}
	\end{center}
\end{figure}

\begin{table}[h!]
  \begin{center}
    \begin{tabular}{|lccc|}
      \hline
		& VP-LBM & FVM  & Experiments \\
      
      $C_{Dave}$ & 1.45   & 1.43 & 1.45 \cite{Wen_2004} \\
      
      $C_{Lrms}$ & 0.84   & 0.82 &                      \\
      
      $St$       & 0.22   & 0.22 & 0.21 \cite{Wen_2001} \\
      \hline
    \end{tabular}
  \end{center}
  \caption{Results obtained for a flow around a cylinder at a Reynolds number of 500}
  \label{Cd_Cl_St_cylinder}
\end{table}

\section{Conclusions}
In this paper, we presented a new approach for modeling fluid flows in the presence of solid obstacles. It consists in coupling the 
volume penalization method with the lattice Boltzmann approach. With this method, the lattice Boltzmann equation was applied on both solid and fluid 
media, and the solid domain was penalized in order to apply the solid velocity in the solid domain.\\
To evaluate this method, several cases were considered. 
First, the VP-LBM method was applied to cases for which an analytical solution is known: a symmetric shear flow, and a cylindrical Couette flow. The 
numerical results obtained with the volume penalization method combined with 
the Single Relaxation Time Lattice Boltzmann 
method on the one hand, and with the Multiple Relaxation Time Lattice Boltzmann method, on the other hand, were compared with 
the analytical solutions, and with results obtained by a combination of the immersed boundary method and the LBM, by Lu et al. \cite{lu_immersed_2012} 
who performed a theoretical analysis of numerical slip velocity that occurs when using the immersed boundary LBM. For the these cases,
we noticed that the VP MRT lattice Boltzmann method enables to reduce the slip velocity in the solid domain, 
in comparison with the VP SRT LBM. For straight boundaries, this error was very close to zero when an optimum value of the relaxation time was used for the SRT LBM, 
and also when an optimum value of the parameter $\Lambda$ was chosen for the MRT LBM. In the case with a curved boundary, the error could not be
eliminated, but it was decreased when the MRT LBM was applied. Then, the flow around a circular cylinder at a Reynolds number equal to
500 was computed. To treat that more complex case, the VP LBM code was parallelized with CUDA, and the computations were performed on Graphics Processing Units.
The results obtained for that case were compared with results found in the literature and with the finite volume method.
A good agreement was highlighted.\\
To conclude with the VP-LBM, it can be noticed that it is  physically meaningful, and easy to implement. Indeed, the computational domain where the 
Navier-Stokes equations (and thus the Lattice Boltzmann equation) are solved consists of the fluid and the solid media, and the volume force that is 
introduced into the Navier-Stokes equations, is applied on the whole solid domain thanks to a mask function, and it considers the obstacle as a porous 
medium. In addition, the convergence of the volume penalization method has been mathematically proved \cite{Angot_1999}. In the context of the Lattice Boltzmann 
method, many researchers apply the immersed boundary method. The immersed boundary method consists in applying in the Navier-Stokes equations a volume force 
located on the surface of the obstacle and in spreading forces on the lattice nodes near the interface  . With this method, the surface of the obstacle must be 
tracked during the computation, and it is less easy to implement 
this method in a highly parallelizable method such as the LBM. Even if the surface of the obstacle is modeled in a staircase  manner with the volume penalization 
method, it enables to calculate flows around complex boundaries, and coupling this method with LBM makes the VP LBM an efficient parallel computing tool.\\

\bibliographystyle{unsrt}
%\bibliography{your_bib_file_without_extension}
%\bibliographystyle{plain}
\bibliography{biblio2}

\end{document}